\RequirePackage{lineno}
\documentclass[pdflatex,aps,prd,twocolumn,nofootinbib,showpacs,superscriptaddress]{revtex4-1}
\usepackage{graphicx}
\usepackage{subfigure}
\usepackage{lipsum}
\newcommand{\nc}[1]{\newcommand{#1}}
\newcommand{\be}{\begin{eqnarray}}
\newcommand{\ee}{\end{eqnarray}}

\nc{\hmu}{\hat{\mu}}
\nc{\bmu}{\bar{\mu}}
\nc{\bchi}{\bar{\chi}}
\def\lsim{\raise0.3ex\hbox{$<$\kern-0.75em\raise-1.1ex\hbox{$\sim$}}}
\def\gsim{\raise0.3ex\hbox{$>$\kern-0.75em\raise-1.1ex\hbox{$\sim$}}}
\usepackage{color}
\usepackage{epsfig}
\usepackage{graphics}
\usepackage{amssymb}
\usepackage{amsmath}
\nc{\fb}{\color{blue}}
\nc{\fr}{\color{red}}

\begin{document}

\title{Skewness and kurtosis of net baryon-number distributions
at small values of the baryon chemical potential
}
\author{
A. Bazavov}
\affiliation{Department of Computational Mathematics, Science and Engineering 
and Department of Physics and Astronomy, Michigan State University, East 
Lansing, MI 48824, USA}
\author{
H.-T. Ding}
\affiliation{Key Laboratory of Quark \& Lepton Physics (MOE) and Institute
of Particle Physics, Central China Normal University, Wuhan 430079, China}
\author{
P. Hegde} 
\affiliation{Center for High Energy Physics, Indian Institute of Science, 
Bangalore 560012, India}
\author{
O. Kaczmarek}
\affiliation{
Fakult\"at f\"ur Physik, Universit\"at Bielefeld, D-33615 Bielefeld,
Germany}
\author{
F. Karsch}
\affiliation{
Fakult\"at f\"ur Physik, Universit\"at Bielefeld, D-33615 Bielefeld,
Germany}
\affiliation{
Physics Department, Brookhaven National Laboratory, Upton, NY 11973, USA}
\author{
E. Laermann} 
\affiliation{
Fakult\"at f\"ur Physik, Universit\"at Bielefeld, D-33615 Bielefeld,
Germany} 
\author{
Swagato Mukherjee} 
\affiliation{
Physics Department, Brookhaven National Laboratory, Upton, NY 11973, USA}
\author{\\ H. Ohno}
\affiliation{
Physics Department, Brookhaven National Laboratory, Upton, NY 11973, USA}
\affiliation{Center for Computational Sciences, University of Tsukuba,
Tsukuba, Ibaraki 305-8577, Japan}
\author{
P. Petreczky}
\affiliation{
Physics Department, Brookhaven National Laboratory, Upton, NY 11973, USA}
\author{E. Rinaldi}
\affiliation{
RIKEN-BNL Research Center, Brookhaven National Laboratory, Upton, NY 11973, USA}
\author{
H. Sandmeyer}
\affiliation{
Fakult\"at f\"ur Physik, Universit\"at Bielefeld, D-33615 Bielefeld,
Germany}
\author{
C. Schmidt} 
\affiliation{
Fakult\"at f\"ur Physik, Universit\"at Bielefeld, D-33615 Bielefeld,
Germany}
\author{
Chris~Schroeder}
\affiliation{Physics Division, Lawrence Livermore National Laboratory, 
Livermore CA 94550, USA}
\author{S. Sharma}
\affiliation{
Physics Department, Brookhaven National Laboratory, Upton, NY 11973, USA}
\author{
W. Soeldner}
\affiliation{
Institut f\"ur Theoretische Physik, Universit\"at Regensburg,
D-93040 Regensburg, Germany \\[2mm]
{\bf (HotQCD Collaboration)}
}
\author{
R.A. Soltz}
\affiliation{Physics Division, Lawrence Livermore National Laboratory, 
Livermore CA 94550, USA}
\author{
P. Steinbrecher}
\affiliation{
Fakult\"at f\"ur Physik, Universit\"at Bielefeld, D-33615 Bielefeld,
Germany}
\affiliation{
Physics Department, Brookhaven National Laboratory, Upton, NY 11973, USA}
\author{
P.M. Vranas}
\affiliation{Physics Division, Lawrence Livermore National Laboratory, 
Livermore CA 94550, USA} 

\date{\today}

\begin{abstract}
We present results for the ratios of mean ($M_B$), variance 
($\sigma_B^2$), skewness ($S_B)$ and kurtosis ($\kappa_B$)
of net baryon-number fluctuations obtained in lattice QCD 
calculations 
with physical values of light and strange quark masses.
Using next-to-leading order Taylor expansions in baryon chemical potential
we find that qualitative features of these ratios
closely resemble the corresponding experimentally measured cumulants
ratios of net proton-number fluctuations for beam energies down to
$\sqrt{s_{_{NN}}} \ge 19.6$~GeV.
We show that the difference in cumulant ratios for the mean net baryon-number,
$M_B/\sigma_B^2=\chi_1^B(T,\mu_B)/\chi_2^B(T,\mu_B)$ and the 
normalized skewness, $S_B\sigma_B=\chi_3^B(T,\mu_B)/\chi_2^B(T,\mu_B)$,
naturally arises in QCD thermodynamics. Moreover, we establish a close
relation between skewness and 
kurtosis ratios, $S_B\sigma_B^3/M_B=\chi_3^B(T,\mu_B)/\chi_1^B(T,\mu_B)$ 
and $\kappa_B\sigma_B^2=\chi_4^B(T,\mu_B)/\chi_2^B(T,\mu_B)$, valid 
at small values of the baryon chemical potential. 
\end{abstract}

\maketitle

\section{Introduction}

Fluctuations of \cite{Hatta:2003wn,Ejiri} and correlations among 
\cite{Koch:2005vg} 
conserved charges of strong interactions have long been considered as 
sensitive observables to explore the structure of the phase
diagram of Quantum Chromodynamics (QCD). These are accessible to 
lattice QCD calculations \cite{Ding:2015ona} and are also the 
most promising observables in the experimental search for the conjectured
critical point \cite{CEP,CP} in the phase diagram of QCD that is currently
underway with the beam energy scan (BES) program at the Relativistic Heavy 
Ion Collider (RHIC) \cite{Luo:2017faz}. The results on net electric charge
\cite{Adamczyk:2014fia,Adare:2015aqk} and net proton-number \cite{STARp08,STARp20,Thader:2016gpa} 
fluctuations obtained from the first BES runs at RHIC have not 
yet provided  conclusive evidence for the existence of a critical point.
However, the data on the skewness and kurtosis
of the distribution of net proton-number fluctuations show an intriguing 
dependence on the beam energy.
The published data on cumulants of net proton-number fluctuations
\cite{STARp08} and, in particular, the preliminary data set on 
net proton-number fluctuations measured in an extended transverse momentum range 
\cite{STARp20,Thader:2016gpa}, show obvious deviations from the thermodynamics 
of a 
hadron resonance gas (HRG). The ratios of even order cumulants, as well
as the ratios of odd order cumulants differ from unity, and different mixed ratios 
formed from even and odd order cumulants are not identical. This may not be 
too surprising as HRG model calculations are not expected to give an accurate 
description of the thermodynamics of strong interaction matter,  
described by QCD. However, these experimental findings  raise the question 
whether the observed pattern seen in net proton-number fluctuations can 
be understood in terms of QCD thermodynamics, which provides
information on net baryon-number fluctuations in equilibrium
\cite{Kitazawa:2012at}, or whether other effects 
such as acceptance cuts, limited efficiencies and rapidity dependence
\cite{Bzdak:2012ab,Ling:2015yau,Bzdak:2013pha,Karsch:2015zna,Bzdak:2017ltv} 
or non-equilibrium effects 
\cite{Berdnikov:1999ph,Kitazawa:2013bta,Mukherjee:2015swa,Mukherjee:2016kyu}, 
are responsible for these differences (for a recent review see
\cite{Asakawa:2015ybt}). 

At large beam energies net proton-number densities are small 
and the baryon chemical potential ($\mu_B$) is close to zero, e.g.
$\mu_B/T\simeq 0.2$ at $\sqrt{s_{_{NN}}}=200$~GeV.
It is, thus, conceivable that low order Taylor expansions are also 
suitable for the description of 
the cumulants of the net baryon-number fluctuations
at the time of freeze-out \cite{Gavai:2010zn}. In fact, 
the calculations of various thermodynamic observables in a Taylor series 
in baryon chemical potential suggest that expansions up to 
next-to-leading order (NLO)  in $\mu_B$ provide good approximations
for these observables for $\mu_B/T\lsim (1.5-2)$ \cite{Bazavov:2017dus}. In
the transition region, characterized by the pseudo-critical temperature
for the chiral transition, $T_c = 154(9)$~MeV \cite{Bazavov:2011nk}, one 
thus may expect to
obtain reliable results up to baryon chemical potentials 
$\mu_B \simeq (225-300)$~MeV within NLO Taylor expansion. 
Comparison of cumulant ratios of net
electric charge fluctuations, measured at 
various beam energies, with lattice QCD results 
\cite{Karsch:2012wm,Bazavov:2012vg,Mukherjee:2013lsa,Borsanyi:2013hza} 
as well as 
HRG model calculations \cite{Adamczyk:2014fia,Cleymans:2005xv,Andronic:2005yp} suggests that 
$\mu_B < 1.4 T$ for $\sqrt{s_{_{NN}}}\ge 19.6$~GeV. NLO 
Taylor expansions
of cumulants of conserved charge fluctuations, thus, should provide
an adequate description of equilibrium thermodynamics of strong
interaction matter for a large part of the beam energies probed with 
the BES at RHIC, $7.7~{\rm GeV}\le \sqrt{s_{_{NN}}}\le 200$~GeV.

The purpose of this paper is to determine, within the framework of
equilibrium thermodynamics of QCD, the dependence of net baryon-number
fluctuations on temperature and baryon chemical potential along
lines in the $T$-$\mu_B$ plane.
We will focus on an analysis of thermodynamic properties of ratios of 
cumulants formed from the first four cumulants of net 
baryon-number fluctuations, i.e.
mean ($M_B$), variance ($\sigma_B^2$), skewness ($S_B$) and kurtosis
($\kappa_B$) of net baryon-number distributions, 
\begin{eqnarray}
R_{12}^B(T,\mu_B)&\equiv& \frac{\chi_1^B(T,\mu_B)}{\chi_2^B(T,\mu_B)} 
\equiv \frac{M_B}{\sigma_B^2}\; , \nonumber \\
R_{31}^B(T,\mu_B)&\equiv& \frac{\chi_3^B(T,\mu_B)}{\chi_1^B(T,\mu_B)} 
\equiv \frac{S_B\sigma_B^3}{M_B}\; , \nonumber \\
R_{42}^B(T,\mu_B)&\equiv& \frac{\chi_4^B(T,\mu_B)}{\chi_2^B(T,\mu_B)} 
\equiv \kappa_B \sigma_B^2 \; .
\label{cumulants}
\end{eqnarray}
Here the $n$-th order cumulants, $\chi_n^B(T,\mu_B)$, are obtained
from partial derivatives of the QCD pressure, $P(T,\mu_B,\mu_Q,\mu_S)$, 
with respect to the baryon chemical potential $\mu_B$. 
Obviously, the ratio $R_{32}^B\equiv S_B\sigma_B$, which also is considered 
frequently as an experimental observable, is simply obtained from the 
above three independent ratios, 
\begin{equation}
R_{32}^B(T,\mu_B) =R_{31}^B R_{12}^B = \frac{\chi_3^B(T,\mu_B)}{\chi_2^B(T,\mu_B)} 
\equiv S_B\sigma_B \; .
\label{R32}
\end{equation}

We will provide lattice QCD results on cumulants of
conserved charge fluctuations in next-to-leading order
Taylor expansions. We will confront these results with experimental data 
on cumulants of net proton-number fluctuations 
($M_P, \sigma_P^2, S_P, \kappa_P$) \cite{STARp08,STARp20}, obtained
by the STAR collaboration during the first BES at RHIC.  
Already at large beam energies, i.e. small values of the baryon
chemical potential, these data significantly deviate from expectations based
on HRG model calculations,
which are commonly used as a baseline for the
analysis of data on particle yields and charge fluctuations in terms
of equilibrium thermodynamics \cite{HRG}. 
In particular data suggest, that
\begin{itemize}
\item 
the ratio $M_P/\sigma_P^2$ is a monotonically
decreasing function of $\sqrt{s_{_{NN}}}$, and
$M_P/\sigma_P^2 > S_P\sigma_P$ in the entire range of $\sqrt{s_{_{NN}}}$
probed so far.
\item
$S_P\sigma_P^3/M_P$ is smaller than unity and 
has a weak but significant dependence on $\sqrt{s_{_{NN}}}$ 
becoming increasingly smaller than unity
with decreasing $\sqrt{s_{_{NN}}}$ or, 
equivalently, with increasing $M_P/\sigma_P^2$.
\item 
For $M_P/\sigma_P^2=0$ or, equivalently, 
for large $\sqrt{s_{_{NN}}}$, 
the relation $S_P\sigma_P^3/M_P \simeq \kappa_P\sigma_P^2$ seems to
hold quite well even though both ratios individually remain smaller
than unity. 
\item
With decreasing $\sqrt{s_{_{NN}}}$ or, equivalently, 
increasing $M_P/\sigma_P^2$,  
the cumulant ratio $ \kappa_P\sigma_P^2$ departs further away from unity 
than the skewness ratio $ S_P\sigma_P^3/M_P$.
It seems that the inequality $ \kappa_P\sigma_P^2 < S_P\sigma_P^3/M_P < 1$
holds
at least for all beam energies $\sqrt{s_{_{NN}}}\ge 19.6$~GeV.
\end{itemize}

We will show here that the QCD calculations of net baryon-number fluctuations
up to NLO in $\mu_B/T$ show all the above qualitative features of the 
cumulants of net proton-number fluctuations.

The paper is organized as follows: 
In Section II we introduce the basic expressions for the NLO expansions 
of cumulants of conserved charge fluctuations. In particular, we 
will derive the formulas needed to describe the variation of
ratios of cumulants on a line in the $T$-$\mu_B$ plane
of the QCD phase diagram. Details of our lattice QCD calculations 
are described in Section III. In Sections IV and V we present 
results on the LO and NLO
Taylor coefficients of cumulant ratios as function of  
$\mu_B$.
We compare these NLO lattice QCD results on net baryon-number fluctuations to 
experimental data on net proton-number fluctuations in Section VI.
Finally, we summarize the QCD results on the next-to-leading order
expansion of cumulant ratios and give some conclusions in Section VII.
Further details on the NLO expansion coefficients are presented 
in an Appendix. 

\section{Taylor expansions of cumulants of conserved charge fluctuations}

\subsection{Expansions at fixed temperature}
For small values of the baryon chemical potential the cumulants of 
net baryon-number $(B)$, net electric charge $(Q)$ or net strangeness
($S$) fluctuations, 
\begin{equation}
\chi_n^X (T,\vec{\mu}) = \frac{\partial^n P/T^4}{\partial(\mu_X/T)^n} \; ,\;
X=B,\ Q,\ S,
\label{chiX}
\end{equation}
with 
$\vec{\mu}=(\mu_B, \mu_Q, \mu_S)$, 
are easily obtained from a Taylor expansion of the QCD pressure. Using
$\hmu_X\equiv \mu_X/T$ the pressure is given by,
\begin{equation}
\frac{P(T,\vec{\mu})}{T^4} =
\sum_{i,j,k=0}^\infty \frac{1}{i!j!k!} \chi_{ijk}^{BQS}(T) \hmu_B^i
\hmu_Q^j \hmu_S^k \; ,
\label{pressure}
\end{equation}
where the expansion coefficients $\chi_{ijk}^{BQS}$ are generalized
susceptibilities,
\begin{equation}
\chi_{ijk}^{BQS} (T) =\left. \frac{\partial^{(i+j+k)} P/T^4}{\partial\hmu_B^i\partial\hmu_Q^j \partial\hmu_S^k}\right|_{\vec{\mu}=0}\; ,\;
\label{gsus}
\end{equation}
which can be evaluated in lattice QCD calculations performed at vanishing
chemical potential. They are functions of the temperature, which we usually 
will not mention explicitly, 
$\chi_{ijk}^{BQS}\equiv \chi_{ijk}^{BQS}(T)$. 
We give the arguments only for the non-expanded cumulants which
are functions of $T$ as well as $\vec{\mu}$, i.e. $\chi_n^X(T,\vec{\mu})$.
In the following we will also adopt the convention to suppress 
subscripts and superscripts of the expansion coefficients whenever a
subscript vanishes, e.g. $\chi_{101}^{BQS} \equiv \chi_{11}^{BS}$ etc.

We will focus on NLO expansions
of the first four cumulants along a line in the space of the three
chemical potentials. This line is fixed by two constraints which relate the
electric charge and strangeness chemical potentials to the baryon
chemical potential \cite{Bazavov:2012vg}. In NLO one may parametrize
them as
\begin{eqnarray}
\hmu_Q (T,\mu_B) &=& q_1(T)\ \hmu_B + q_3(T)\ \hmu_B^3 +{\cal O}(\hmu_B^5) 
\;  , \; \nonumber \\
\hmu_S (T,\mu_B) &=& s_1(T)\ \hmu_B + s_3(T)\ \hmu_B^3 +{\cal O}(\hmu_B^5)  
\; .
\label{mu_expansion}
\end{eqnarray}
In applications to heavy ion collisions it is appropriate to determine the 
coefficients $q_i,  s_i$ from constraints demanding overall vanishing
net strangeness density, 
\begin{equation}
n_S\equiv \chi_1^S(T,\vec{\mu}) =0 \; ,
\label{strange0}
\end{equation}
and a fixed relation 
between net baryon-number and net electric charge densities, 
\begin{equation}  
\frac{n_Q}{n_B}\equiv \frac{\chi_1^Q(T,\vec{\mu})}{\chi_1^B(T,\vec{\mu})}
= \frac{N_P}{N_P+N_N} \; . 
\label{QBratio}
\end{equation}
Here the last equality relates the ratio of the number densities 
to the proton ($N_P$) and neutron ($N_N$) numbers of the incident nuclei
in heavy ion collision experiments.
In the case of gold or uranium nuclei, which are frequently used
in heavy ion collision experiments, setting
$n_Q/n_B= 0.4$ is appropriate \cite{Bazavov:2012vg}. The isospin symmetric case 
obviously corresponds to $n_Q/n_B=1/2$. In that case $\mu_Q=0$ and thus $q_i=0$ at all
orders. Explicit expressions for $q_i,  s_i$ have been given in 
Appendix~B of Ref.~\cite{Bazavov:2017dus}.

We will discuss Taylor expansions for the ratios of cumulants introduced
in Eqs.~\ref{cumulants} and \ref{R32}. Using the parametrization of 
$\mu_Q$ and $\mu_S$ given in Eq.~\ref{mu_expansion}, we 
may write these expansions in terms of $\hmu_B$ up to NLO,
\begin{eqnarray}
R_{12}^B(T,\mu_B) &=& r_{12}^{B,1} \hmu_B + r_{12}^{B,3} \hmu_B^3 \; ,
\label{R12B} \\
R_{31}^B(T,\mu_B) &=& r_{31}^{B,0} + r_{31}^{B,2} \hmu_B^2 \; ,
\label{R31B} \\
R_{42}^B(T,\mu_B) &=& r_{42}^{B,0} + r_{42}^{B,2} \hmu_B^2 \; .
\label{R42B} 
\end{eqnarray}
Here the expansion coefficients $r_{nm}^{B,k}$ are functions
of temperature and the Taylor expansion coefficients $q_i, s_i$ of the
constraint chemical potentials $\mu_Q, \mu_S$. The superscript $k$ labels the 
order of the expansion in terms of $\mu_B$. Explicit expressions
for the expansion coefficients $r_{nm}^{B,k}$ in terms of the
generalized susceptibilities, introduced
in Eq.~\ref{gsus}, are given in the Appendix.

\subsection{Expansions along lines \boldmath$T(\mu_B)$ in the $T$-$\mu_B$ plane}

It is our goal to compare cumulant ratios measured in heavy ion 
experiments at different beam energies, $\sqrt{s_{_{NN}}}$, with
lattice QCD calculations of such ratios.
As the beam energy is varied also
the thermal conditions under which particles ``freeze-out" change. This is 
commonly characterized by a pair of freeze-out parameters $(T_f,\mu_B)$. They 
map out a line, $T_f(\mu_B)$, in the QCD phase diagram. When comparing
the Taylor expanded cumulant ratios, discussed in the previous subsection,
with experimental data we thus also need to take into account that
the freeze-out temperature varies with increasing $\mu_B$. 
At large beam energies it is
appropriate to parametrize the freeze-out line as a 
polynomial in $\mu_B^2$ \cite{Cleymans:2005xv}\footnote{Such a 
parametrization is commonly used
when comparing experimental data on particle yields with
statistical hadronization models (HRG models).
An alternative parametrization, used in
Ref.~\cite{Andronic:2005yp}, also provides a good description
of the experimental data but does not have a polynomial behavior for
small $\mu_B$. It starts out with exponentially small corrections to
the freeze-out temperature at vanishing $\mu_B$.}.

In the context of Taylor expansions for bulk thermodynamic
observables also `lines of constant physics' \cite{Bazavov:2017dus} as well 
as the pseudo-critical
line for the QCD transition \cite{Kaczmarek:2011zz,Bonati:2015bha,Bellwied:2015rza,Cea:2015cya} are generally given as 
polynomials in $\mu_B^2$. 
We thus will consider the behavior of
cumulants of conserved charge fluctuations on lines in
the $T$-$\mu_B$ plane
that are parametrized as
\begin{equation}
T_f(\mu_B) = T_{0} \left( 1 - \kappa_2^f \bmu_B^2 +{\cal O}(\bmu_B^4)
\right) \; ,
\label{Tf}
\end{equation}
with $\bmu_B\equiv \mu_B/T_{0}$. 
As we will exploit only NLO expansions for cumulants
it suffices to know this parametrization up to ${\cal O}(\mu_B^2)$.

Taking into account this temperature variation requires an additional
expansion of the ratios $R_{nm}^B$ 
in $T$. On a line $T_f(\mu_B)$
the Taylor expansion in $T$ then generates additional terms that
are of order $\mu_B^2$. I.e. the LO expansion coefficients of cumulant ratios
remain unchanged, while the NLO expansion coefficients, $r_{nm}^{B,k}(T)$, 
receive an additional contribution from the variation of 
cumulant ratios with temperature along a line in the $T$-$\mu_B$ plane,

\begin{eqnarray}
r_{nm}^{B,k} &\rightarrow& r_{nm,f}^{B,k} \equiv r_{nm}^{B,k}(T_{0}) -\kappa_2^f T_{0} 
\left. \frac{{\rm d} r_{nm}^{B,k-2}}{ {\rm d} T}\right|_{T=T_{0}} 
\label{kappa_contribution}
\end{eqnarray}
with $k=2$ or $3$. 
With this the three cumulant ratios introduced in Eq.~\ref{cumulants} become
\begin{eqnarray}
R_{12}^B(T_f(\mu_B),\mu_B) &=& r_{12}^{B,1} \bmu_B +
r_{12,f}^{B,3}\ \bmu_B^3  \; , 
\label{Rc1} \\
R_{31}^B(T_f(\mu_B),\mu_B) &=& r_{31}^{B,0}
+ r_{31,f}^{B,2}\ \bmu_B^2 \; ,
\label{Rc2} \\
R_{42}^B(T_f(\mu_B),\mu_B) &=& r_{42}^{B,0}
+ r_{42,f}^{B,2}\ \bmu_B^2  \; .
\label{Rc3}
\end{eqnarray}
Here all expansion coefficients $r_{nm}^{B,k}$ and $r_{nm,f}^{B,k}$ 
are evaluated at $\mu_B=0$ and at the temperature $T(\mu_B=0)\equiv T_{0}$. 

In the following sections we will present lattice QCD results for the 
expansion coefficients appearing in Eqs.~\ref{Rc1}-\ref{Rc3}.
In particular, as done before in an analysis of ratios of variances of net
electric charge and net baryon-number fluctuations \cite{Bazavov:2015zja} we 
will make use of the fact that $r_{12}^{B,1}$ is positive for all
values of the temperature. At least to leading order in
$\mu_B$ the ratio $M_B/\sigma_B^2$ thus is a monotonically rising function
of $\mu_B$. We may use this to eliminate the baryon
chemical potential $\mu_B$ in the NLO relations
for $R_{31}^B$ and $R_{42}^B$ in favor of the mean net baryon-number ratio, 
$R_{12}^B\equiv M_B/\sigma_B^2$, i.e.
\begin{equation}
\hmu_B = \frac{1}{r_{12}^{B,1}}R_{12}^B +{\cal O}((R_{12}^B)^3) \; .
\label{R12muB}
\end{equation}
With this we obtain at NLO
\begin{eqnarray}
R_{31}^B(T,R_{12}^B) &=& r_{31}^{B,0}
+ \frac{r_{31,f}^{B,2}}{ \left(r_{12}^{B,1}\right)^2}
\left( R_{12}^B\right)^2  \; ,
\label{R31vsR12} \\
R_{42}^B(T,R_{12}^B) &=& r_{42}^{B,0}
+ \frac{r_{42,f}^{B,2}}{\left(r_{12}^{B,1}\right)^2}
\left( R_{12}^B\right)^2 \; .
\label{RnmvsR12}
\end{eqnarray}
Using Eq.~\ref{R32} we easily obtain from Eq.~\ref{R31vsR12} also the NLO 
expansion for the ratio $R_{32}^B$,
\begin{eqnarray}
R_{32}^B(T,R_{12}^B) &=& r_{31}^{B,0} R_{12}^B
+ \frac{r_{31,f}^{B,2}}{ \left(r_{12}^{B,1}\right)^2}
\left( R_{12}^B\right)^3  \; .
\label{R32vsR12}
\end{eqnarray}
Considering expansions of higher order cumulant ratios in
terms of the lowest order ratio $R_{12}^B$ rather than expansions
in $\hmu_B$ has the advantage that we can compare the QCD results
directly to experimental observables without the need of first
determining a chemical potential from the data. A trivial consequence is,
that at LO the slope of the expansion of $R_{32}^B$ in terms of $R_{12}^B$
is identical to the intercept of $R_{31}^B$ at $\mu_B=0$.

We note that in the low temperature HRG limit $R_{12}^B\simeq \tanh \hmu_B$ which
can be inverted for all $\hmu_B$. However, in the vicinity of a possible 
critical 
point in the $T$-$\mu_B$ plane $R_{12}^B$ will no longer be a 
monotonic function of $\mu_B$ as $\sigma_B^2$ will diverge at 
a critical point while $M_B$ stays finite. In the parameter
range probed experimentally so far, no indication for such a 
non-monotonic behavior of $R_{12}^B$ has been observed.

\section{Lattice QCD calculations}

The main results presented in the following are based on lattice
QCD calculations performed in the temperature range 
$135~{\rm MeV} \lsim T \lsim 175~{\rm MeV}$. In this temperature interval
our analysis is based on calculations performed with a strange quark
mass tuned to its physical value and degenerate light quarks with a 
mass $m_l/m_s=1/27$. In the continuum limit this light quark mass
corresponds to a pion mass of about 140~MeV. For completeness and 
in order to give a feeling for the trends in the temperature 
dependence of various observables we added a few data at higher $T$-values 
that have been obtained from calculations with a somewhat larger
quark mass ratio, $m_l/m_s=1/20$. In the continuum limit this quark 
mass ratio corresponds to a pion mass of about 160~MeV.

\begin{table}[t]
\begin{center}
\vspace{0.3cm}
\begin{tabular}{|c|c||c|c||c|c|}
\hline 
\multicolumn{2}{|c||}{$N_\tau=8$}&\multicolumn{2}{|c||}{$N_\tau=12$}&\multicolumn{2}{|c|}{$N_\tau=16$} \\
\hline
 T[MeV] & \#conf.&T[MeV] & \#conf.&T[MeV] & \#conf. \\
\hline
134.64&456,070&134.94&39,380&-&~ \\
140.45&626,790&140.44&61,610&-&~ \\
144.95&684,200&144.97&69,910&144.94&2,980 \\
151.00&362,200&151.10&45,900&151.04&8,080 \\
156.78&513,130&157.13&30,100&156.92&4,850 \\
162.25&247,040&161.94&32,810&162.10&3,010 \\
165.98&283,640&165.91&64,820&166.03&2,510 \\
171.02&139,980&170.77&40,870&170.98&2,430 \\
175.64&137,500&175.77&39,040&-&~ \\
\hline
\end{tabular}
\end{center}
\caption{Number of gauge field configurations on lattices of size 
$32^3\times 8$, $48^3\times 12$ and $64^3\times 16$ used
in the analysis of up to $6^{th}$ order Taylor expansion coefficients.
The values of the gauge coupling as well as the strange and light quark mass 
parameter at these temperature values are taken from \cite{Bazavov:2017dus},
where also details on the statistics available on the $24^3\times 6$
lattices are given. 
}
\label{tab:statistics}
\end{table}

The parameter choices, lattice sizes, quark masses as well as the 
determination of the temperature scale from zero temperature observables,
are identical to the set-up used previously in our calculation of the
equation of state at vanishing chemical potential \cite{Bazavov} and 
the calculation of the equation of state of (2+1)-flavor QCD at non-zero
baryon chemical potential in $6^{th}$ 
order Taylor series \cite{Bazavov:2017dus}. 

Our calculations are performed on lattices of size $N_\sigma^3\times N_\tau$
with $N_\sigma=4N_\tau$ and $N_\tau=6,8,12,16$.
Compared to earlier calculations \cite{Bazavov:2017dus} we have
increased the statistics on the $32^3\times 8$ and $48^3\times 12$
lattices by about a factor four and added new calculations on lattices
of size $64^3\times 16$. 

Our main conclusions on 
the behavior of NLO expansion coefficients are based on calculations
performed on lattices of size $32^3\times 8$, where we generated up to
$700,000$ gauge field configurations using the Rational Hybrid
Monte Carlo (RHMC) algorithm. We generated up to 7 million RHMC trajectories
of unit length and saved gauge field configurations after every $10^{th}$
trajectory. Our current statistics is summarized in Table~\ref{tab:statistics}.

Up to $6^{th}$ order cumulants have been calculated on these
data sets. Due to the large number of gauge field configurations needed
for an analysis of the expansion coefficients of kurtosis and skewness
ratios, our main results on NLO expansion coefficients for these observables
is based on calculations for 
a single lattice size only, i.e. they are not yet continuum
extrapolated, although cut-off effects are expected to be significantly
smaller in theses observables than our current statistical errors.


For the LO observables, continuum extrapolations were done using global spline fits
to data from all 4 lattice spacings following the procedure described in
\cite{Bazavov:2015zja,Bazavov:2017dus}. We allow for $1/N_\tau^2$ dependence of the
spline coefficients, and also vary the locations of the spline knots to minimize the
$\chi^2$ of the global fits. For the current analysis we found it is sufficient to use
spline interpolations with quartic polynomials and 3 knots whose location is allowed
to vary in the fit range. Fits were done for many bootstrap samples drawn from the
Gaussian errors of data points, and final results were obtained from mean values and
standard deviations of the bootstrapped fit results, weighted by the quality of the fits 
given by the Akaike information criteria. For the NLO observables, we have lattice data 
only for 2 lattice spacings corresponding to $N_\tau =6$ and $8$, and could not carry out
proper continuum extrapolations. For these cases, we provide 'continuum estimates'
following exactly the same continuum extrapolation procedure described above, but
only using global spline fits to the data from $N_\tau =6$ and $8$ lattices

In the following three sections we will present results on the various
LO and NLO expansion coefficients entering in Eqs.~\ref{R12muB}-\ref{R32vsR12}.

\begin{figure*}[htb]
\begin{center}
\includegraphics[width=78mm]{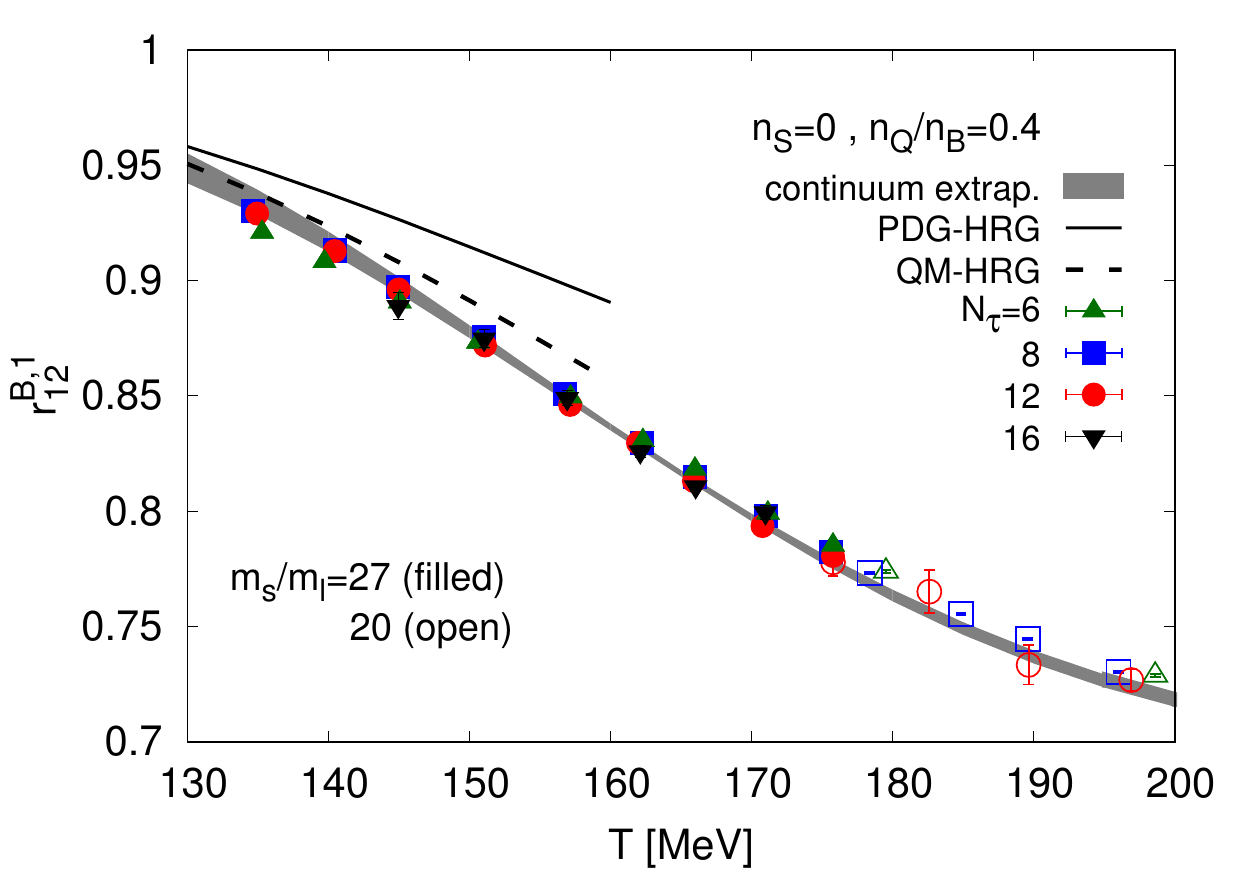}
\includegraphics[width=78mm]{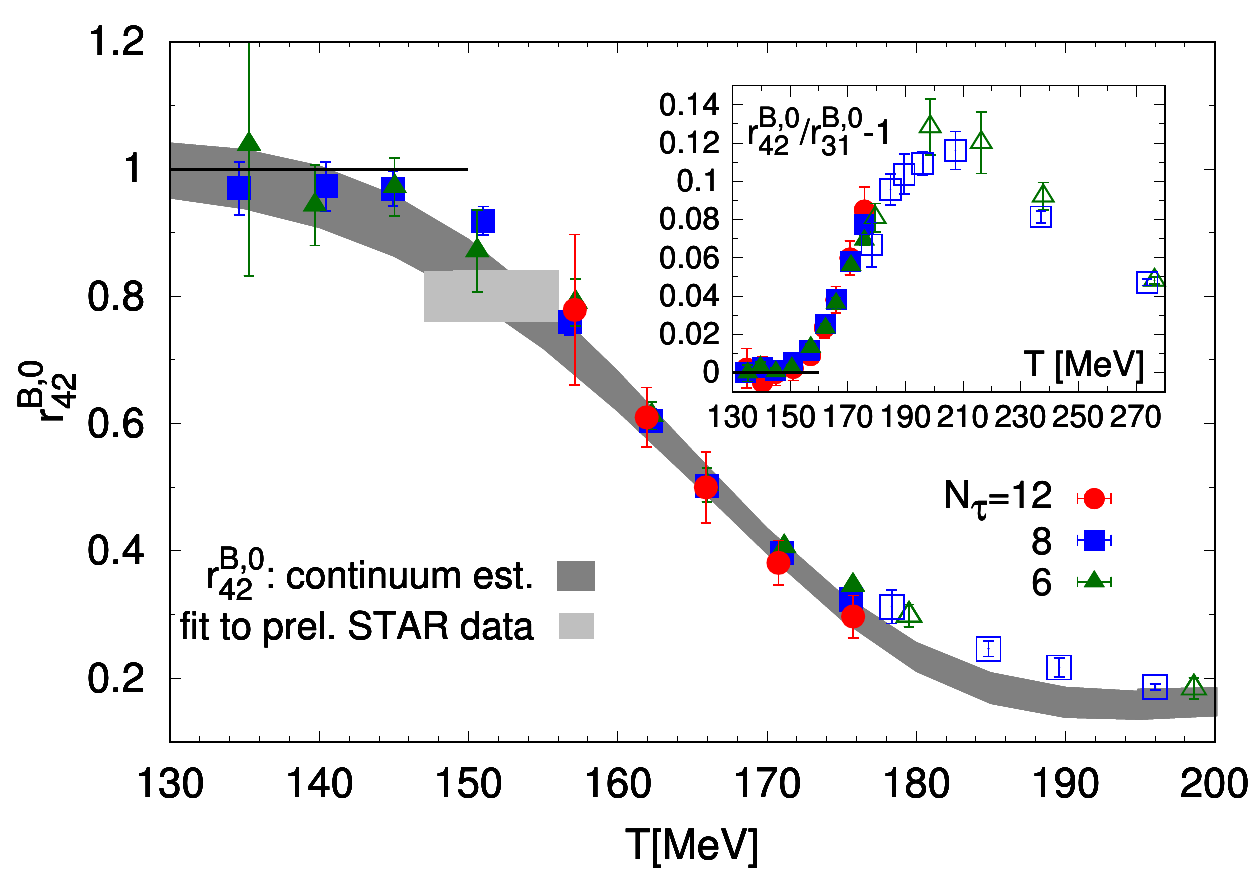}
\caption{
The leading order expansion coefficients of the cumulant ratios
$R_{12}^B$ (left) and $R_{42}^B$ (right)
versus temperature calculated on
lattices with temporal extent $N_\tau$, and spatial sizes $N_\sigma =4 N_\tau$.
The inset in the right hand figure shows the difference between the 
leading order results for the kurtosis ratio $R_{42}^B$ and the 
skewness ratio $R_{31}^B$ normalized to the latter. All expansion coefficients 
have been calculated
for strangeness neutral systems, Eq.~\ref{strange0}, with an electric charge
to baryon number ratio $r=0.4$ (Eq.~\ref{QBratio}).
The grey bands give the continuum extrapolated result for $r_{12}^{B,1}$
and, in the case of $r_{42}^{B,0}$, an estimate for the continuum result.
In the right hand figure we also
show results from a fit to the preliminary STAR data for the corresponding
net proton-number fluctuations discussed in Section~\ref{sec:compare}.
See text for a discussion of the two HRG curves shown in the left hand figure.
}
\label{fig:R12X1}
\end{center}
\end{figure*}

\section{Leading order expansion of cumulant ratios}

The leading order expansion coefficients for the ratios $R_{nm}^B$
defined in Eq.~\ref{cumulants} are given in the Appendix. We can write
them as
\begin{eqnarray}
r_{12}^{B,1} &=& 1+ s_1 \frac{\chi_{11}^{BS}}{\chi_2^B} +q_1 \frac{\chi_{11}^{BQ}}{\chi_2^B} \; ,
\label{r12B1} \\
r_{42}^{B,0} &=& \frac{\chi_4^B}{\chi_2^B} \; ,
\label{r42B0} \\
r_{31}^{B,0} &=& r_{42}^{B,0}
\frac{1+ s_1 \frac{\chi_{31}^{BS}}{\chi_4^B} +q_1 \frac{\chi_{31}^{BQ}}{\chi_4^B}}{1+ s_1 \frac{\chi_{11}^{BS}}{\chi_2^B} +q_1 \frac{\chi_{11}^{BQ}}{\chi_2^B}}
\label{r31B0}
\end{eqnarray}
This makes it apparent that the LO coefficients are particularly simple
for $\mu_S=\mu_Q=0$. In that case one has $r_{12}^{B,1} =1$ and
$r_{31}^{B,0} = r_{42}^{B,0}$. 
In the strangeness neutral case with fixed $n_Q/n_B=0.4$ the 
contribution from a non-vanishing electric charge chemical potential 
is small. The dominant correction arises from a non-zero strangeness
chemical potential needed to insure strangeness neutrality 
\cite{Bazavov:2014xya}. As  $s_1>0$ and $\chi_{11}^{BS}/\chi_2^B<0$
we thus expect to find $r_{12}^{B,1}<1$. This is also the case at low
temperature for a HRG. 

In Fig.~\ref{fig:R12X1} we show results for the leading order 
expansion coefficients of the ratios $R_{12}^B$, $R_{31}^B$ and
$R_{42}^B$, respectively. The left hand figure shows
the LO expansion coefficient $r_{12}^{B,1}$. This is an update
on results obtained previously in \cite{Bazavov:2015zja} from
calculations with much lower statistics.
The right hand part of the figure shows 
the LO result $r_{42}^{B,0}$.  Also shown as an inset in 
this figure is the difference between the leading order
results for $r_{42}^{B,0}$ and $r_{31}^{B,0}$ normalized to the latter.  
The LO results for the cumulant ratios shown in this figure have been 
obtained for a strangeness neutral system, $n_S=0$, with electric charge to 
baryon number ratio $n_Q/n_B =0.4$. 

Let us first discuss the leading order results for the ratio 
$R_{12}^B(T,\mu_B)$. Here results from calculations on lattices
with temporal extent $N_\tau=6$ to $16$ exist. They show rather small 
cut-off dependence, which is known also from our calculations
of the Taylor expansion coefficients of the equation of state. The results
have been extrapolated to the continuum limit using spline interpolations
as described in Ref.~\cite{Bazavov}.
Obviously $r_{12}^{B,1}$ approaches the HRG value from
below at all values of the temperature. As has been observed
previously in calculations of cumulants that are sensitive to the 
strange baryon sector of hadron resonance gas models
\cite{Bazavov:2014xya}, we find that a HRG model, which includes 
additional strange baryons (QM-HRG) provides a better description of the 
Taylor expansion coefficients than a HRG model
based only on experimentally well established
resonances listed in the Particle Data Tables (PDG-HRG) \cite{PDG}.

Similarly, the LO expansion coefficients of the ratios
$R_{31}^B\equiv S_B\sigma^3_B/M_B = r_{31}^{B,0} +{\cal O}(\mu_B^2)$ and
$R_{42}^B\equiv \kappa_B\sigma^2_B = r_{42}^{B,0} +{\cal O}(\mu_B^2)$,
shown in Fig.~\ref{fig:R12X1}~(right),
seem to approach the HRG model value from below.
At least for $T>150$~MeV these ratios are smaller than unity. As a consequence
we find to LO in $\mu_B$, or equivalently to LO in $R_{12}^B$, 
that 
\begin{equation}
R_{32}^B \equiv R_{31}^B R_{12}^B < R_{12}^B +{\cal O}((R_{12}^B)^3) \; .
\end{equation}

At least for $T>150$~MeV ratios of cumulants thus obey the inequality
$M_B/\sigma_B^2 > S_B\sigma_B$ or equivalently
$R_{31}^B \equiv S_B\sigma_B^3/M_B <1$. 
This clearly is different from HRG model calculations with point-like, 
non-interacting hadrons, where the exact relations,
$M_B/\sigma_B^2 = S_B\sigma_B$  and $ S_B\sigma_B^3/M_B =1$, hold at any
order in $\mu_B$, irrespective of the details of the hadron spectrum used
in that calculation. 

From the LO expressions given in 
Eqs.~\ref{r42B0} and \ref{r31B0} it is obvious that to leading order  
the ratios $R_{31}^B$ and $R_{42}^B$ 
will also be identical in the case of vanishing strangeness
and electric charge chemical potentials, although their values need not
be unity. Fig.~\ref{fig:R12X1}~(right) shows that the LO
coefficient $r_{42}^{B,0}$ starts to deviate from unity significantly
for $T>150$~MeV.
Nonetheless, as can be seen from the inset in 
Fig.~\ref{fig:R12X1}~(right) the difference of the LO expansion coefficients,
$r_{42}^{B,0}-r_{31}^{B,0}$, stays small also in the strangeness neutral
case with $n_Q/n_B=0.4$. The maximal difference is reached at $T\simeq 200$~MeV
where it amounts to about 12\% of $r_{31}^{B,0}$.
However, in the crossover region, $145~{\rm MeV} < T < 165~{\rm MeV}$,
which also is the temperature range of interest for comparison with 
experimental data, this difference never exceeds more than 4\% of 
$r_{31}^{B,0}$. The experimental observation that
$S_P\sigma_P^3/M_P$ and $\kappa_P\sigma_P^2$ tend to agree at large
$\sqrt{s_{_{NN}}}$, although they differ from unity, thus is in accordance 
with the QCD result,
\begin{equation}
S_B\sigma_B^3/M_B \simeq \kappa_B\sigma_B^2 \;\;\; {\rm for} 
\;\;\; R_{12}^B\rightarrow 0
\; .
\label{limit}
\end{equation}

\section{Next-to-Leading order expansions of cumulant ratios}

The NLO corrections in the series expansion of ratios $R_{nm}^B$
at fixed temperature as well as on lines in the 
$T$-$\mu_B$ plane have been introduced in Eqs.~\ref{R12B}-\ref{R42B} 
and in Eqs.~\ref{Rc1}-\ref{Rc3}, respectively.
We will derive the NLO expansion coefficients in the following and 
show results for strangeness neutral systems
with an electric charge to baryon number ratio $n_Q/n_B=0.4$.
However, for the discussion presented in this section we will 
also use the simpler expressions obtained for the case
of vanishing strangeness and electric charge chemical potentials. 
In this case the information contained in the NLO expansion
coefficients is much more transparent and, as we will see,
they show the same qualitative features and furthermore yield similar 
quantitative results.

The NLO expansions for cumulants and the resulting expansions
of cumulant ratios for arbitrary values of the chemical potentials
$\vec{\mu}=(\mu_B,\mu_Q,\mu_S)$ are given in the Appendix. From
these one easily obtains
the NLO expansion coefficients $r_{12}^{B,3}$, $r_{31}^{B,2}$ 
and $r_{42}^{B,2}$ for $\mu_Q=\mu_S=0$ by evaluating these
expressions for $s_i=q_i=0$ for $i=1,\ 3$.
This yields for the ratio of NLO and LO expansion coefficients,
\begin{eqnarray}
\frac{r_{12}^{B,3}}{r_{12}^{B,1}} &=& -\frac{1}{3} \frac{\chi_4^B}{\chi_2^B} \; ,
\label{R-R12B30} \\
\frac{r_{31}^{B,2}}{r_{31}^{B,0}} &=& 
\frac{1}{6} \left( \frac{\chi_6^B}{\chi_4^B} - 
\frac{\chi_4^B}{\chi_2^B} \right) \; ,
\label{R-R31B20} \\
\frac{r_{42}^{B,2}}{r_{42}^{B,0}} &=& 3 \frac{r_{31}^{B,2}}{r_{31}^{B,0}} \; .
\label{R-R42B20}
\end{eqnarray}
As the quadratic and quartic cumulants of net baryon-number fluctuations
are positive for all values of the temperature \cite{Bazavov:2017dus}, 
the NLO expansion coefficient 
of $R_{12}^B=M_B/\sigma_B^2$  is negative for all $T$. The NLO
expansion coefficient of $R_{31}^B=S_B\sigma_B^3/M_B$ is negative as 
long as $\chi_6^B/\chi_4^B <\chi_4^B/\chi_2^B$. As known from the Taylor
expansion of the equation of state (Fig.~13 of Ref.~\cite{Bazavov:2017dus})
this is the case at least for $T\gsim 155$~MeV. Furthermore, 
Eq.~\ref{R-R42B20} explicitly states that the NLO correction to the 
kurtosis ratio $R_{42}^B$ is three times larger than that for the 
skewness ratio $R_{31}^B$ for all $T$ as long as $\mu_Q=\mu_S=0$.

Using Eqs.~\ref{R-R12B30} and \ref{R-R31B20} it also is straightforward
to obtain the NLO expansion coefficient of $R_{32}^B\equiv S_B\sigma_B$,
\begin{equation}
\frac{r_{32}^{B,3}}{r_{32}^{B,1}} = \frac{r_{31}^{B,2}}{r_{31}^{B,0}} +
\frac{r_{12}^{B,3}}{r_{12}^{B,1}}  =
\frac{1}{6} \frac{\chi_6^B}{\chi_4^B} - 
\frac{1}{2} \frac{\chi_4^B}{\chi_2^B} 
\; ,
\label{R-R32B30}
\end{equation}
which also is negative at least for $T\gsim 155$~MeV (see Fig.~13 of 
Ref.~\cite{Bazavov:2017dus}).

\begin{figure*}[t]
\begin{center}
\includegraphics[width=78mm]{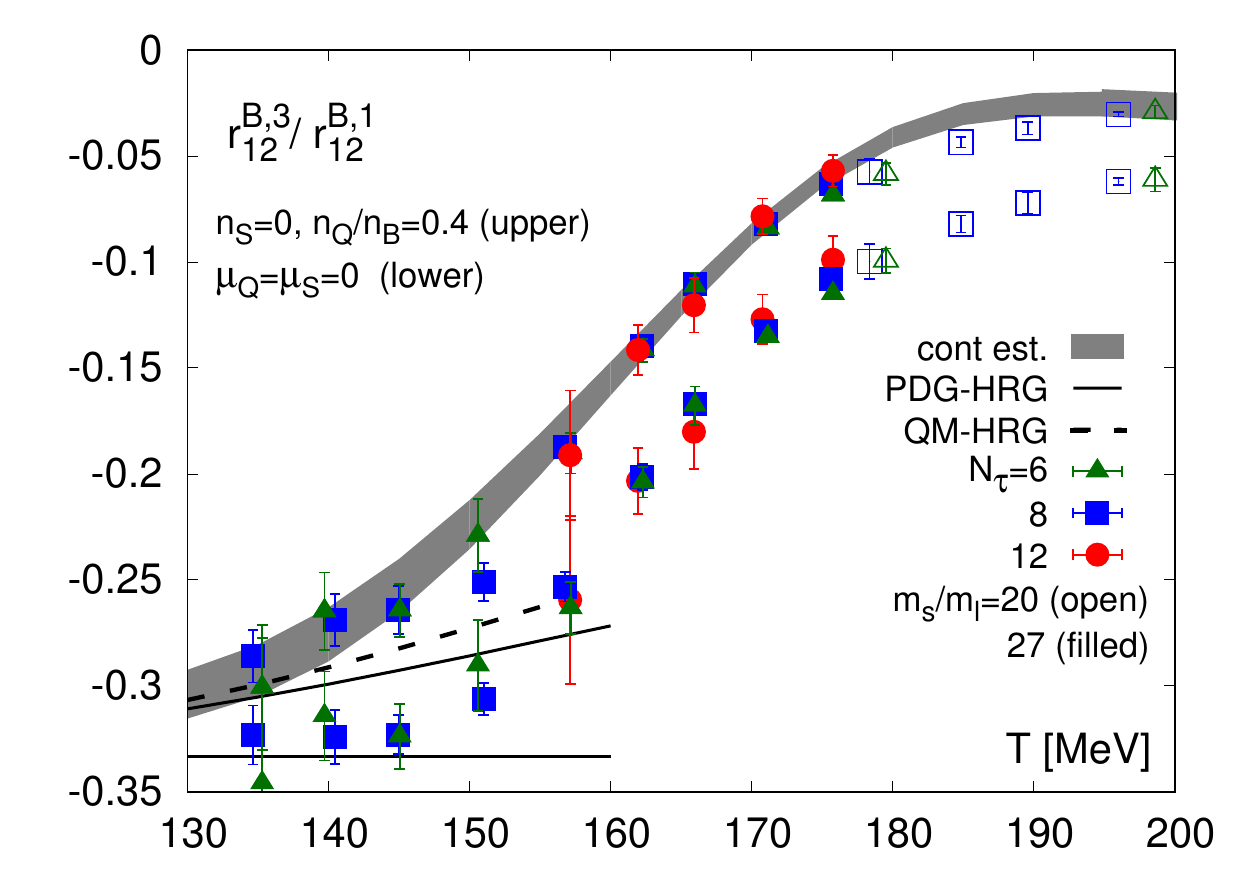}
\includegraphics[width=78mm]{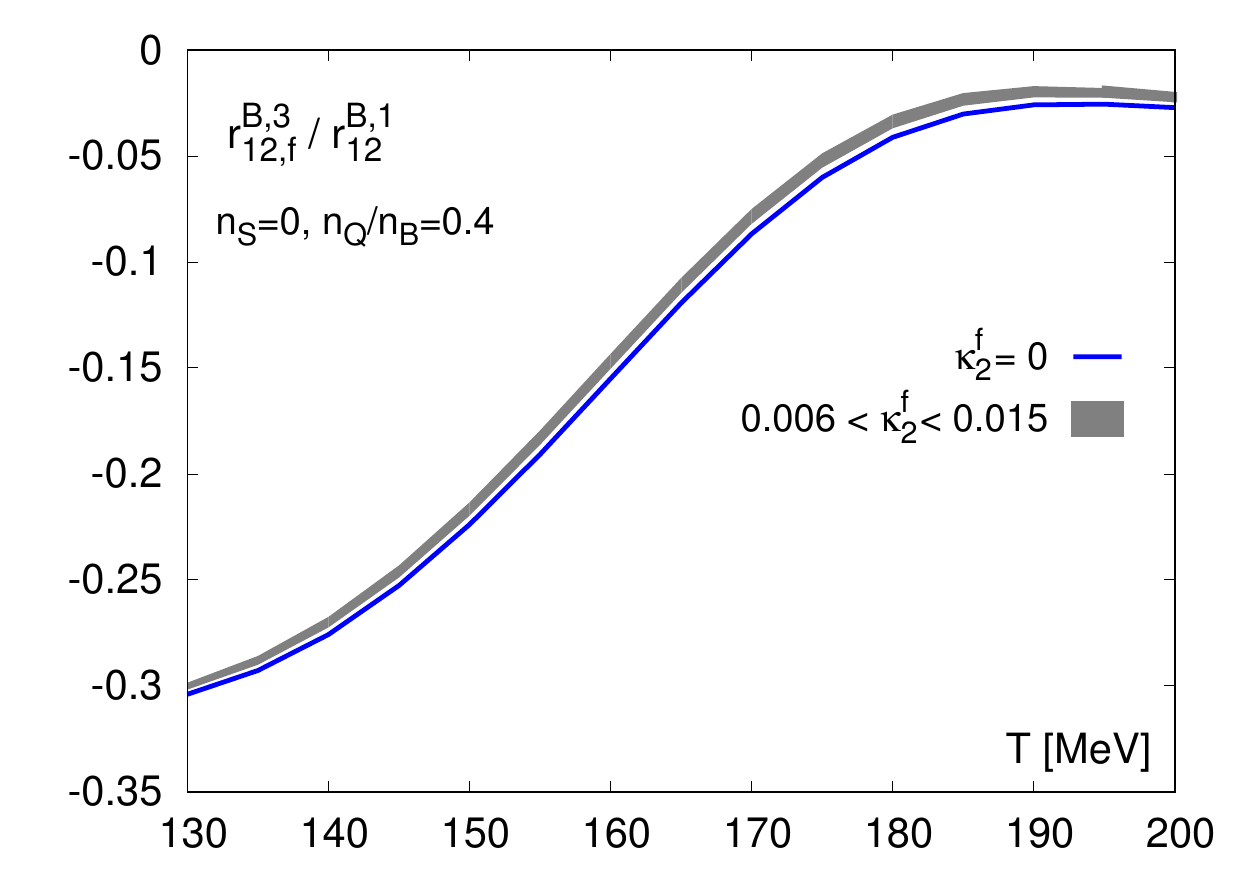}
\caption{Ratio of NLO and LO expansion coefficients of the
cumulant ratio $R_{12}^B\equiv M_B/\sigma_B^2$ versus temperature.
The left hand figure shows results for the NLO expansion coefficient
evaluated at fixed temperature while the right hand figure gives
results for the NLO expansion coefficient on a line in the $T$-$\mu_B$ plane
defined in Eq.~\ref{Tf}.
The lower data set in the left hand figure correspond to the
case $\mu_Q=\mu_S=0$ and the upper data set correspond to
the strangeness neutral case with $n_Q/n_B=0.4$.
See section IV for a discussion of the HRG model curves shown in the
left hand figure.
The right hand figure shows fits to the
ratio $r_{12,f}^{B,3}/r_{12}^{B,1}$ in the strangeness neutral case
for expansion coefficients at fixed temperature, i.e. for $\kappa_f=0$
(lower line) and
on lines, $T_f(\mu_B)$, with curvature coefficients in the range
$0.006\le \kappa_2^f\le 0.015$. For clarity no error band is shown
in this figure.
}
\label{fig:R12B3}
\end{center}
\end{figure*}

\subsection{NLO expansion coefficients of \boldmath$R_{12}^B$}
In Fig.~\ref{fig:R12B3} we show results for the ratio of NLO and LO 
expansion coefficients of $R_{12}^B$.
The left hand figure shows the ratio of expansion
coefficients for a Taylor series evaluated at fixed temperature
for the two cases (i) $\mu_S=\mu_Q=0$
and (ii) $n_S=0$, $n_Q/n_B=0.4$. 
It is obvious that the simpler case (i) is
qualitatively similar to the strangeness neutral case (ii).
However, in the latter case the ratio of NLO and LO expansion coefficients is
systematically smaller in magnitude.

In order to judge the importance of additional contributions
to NLO expansion coefficients, that arise from the variation of $T$
along a line $T_f(\mu_B)$ in the $T$-$\mu_B$ plane, we use the parametrization
given in Eq.~\ref{Tf}. We are particularly
interested in a line that characterizes the change of freeze-out conditions
that results from changes of the beam energy in heavy ion collisions
experiments.
Of course, such a line eventually needs to be determined from the experimental
data. However, it has been suggested  \cite{Cleymans:2005xv,Castorina:2014fna}
that hadronic freeze-out in heavy ion collisions may take place along a line 
on which some thermodynamic observables stay constant as functions
of $(T,\mu_B)$. 
Such ``lines of constant physics'' have been determined
from the Taylor expansions of bulk thermodynamic observables
\cite{Bazavov:2017dus} up to ${\cal O} (\mu_B^4)$.
For the purpose of our current NLO analysis it suffices to use
information from these expansions that defines the lines
$T_f(\mu_B)$ up to ${\cal O}(\mu_B^2)$.
It turns out that
lines of constant pressure, energy density or entropy density
describe similar trajectories in the $T$-$\mu_B$ plane. At NLO 
such lines are controlled by the curvature coefficient $\kappa_2^P$
(pressure), $\kappa_2^\epsilon$ (energy density) or
$\kappa_2^s$ (entropy density), respectively. In the crossover region, 
$T_c=(154\pm 9)$~MeV, we find that these curvature coefficients vary in the 
range\footnote{The temperature dependence of these curvature coefficients 
for the three different bulk thermodynamic observables is shown in
Fig.~14 of Ref.~\cite{Bazavov:2017dus}.},
\begin{equation}
0.006 \le \kappa_2^f \le 0.012 \; ,\; f=P, \epsilon, s\;  .
\label{kappa2}
\end{equation}
For baryon chemical potentials $\mu_B/T\le 1.5$ the temperature variation
on a line $T_f(\mu_B)$ with $\kappa_2^f \le 0.012$ thus is less than 3\%
of the $T$-value at $\mu_B=0$.
The $\mu_B$-dependence of the chiral crossover transition \cite{Kaczmarek:2011zz,Bonati:2015bha,Bellwied:2015rza,Cea:2015cya}
is similar in magnitude. This range of curvature coefficients also is
consistent with the bound on $\kappa_2^f$ extracted in \cite{Bazavov:2015zja}
by comparing experimental data for $M_P/\sigma_P^2$ and the corresponding
electric charge ratio $M_Q/\sigma_Q^2$ with a NLO lattice QCD calculation.

The right hand part of Fig.~\ref{fig:R12B3} shows the influence of a 
non-vanishing curvature coefficient, $\kappa_2^f\le 0.015$, on the NLO
expansion coefficients for $R_{12}^B\equiv M_B/\sigma_B^2$. 
As can be seen the modification is small, leading at most to a 10\% change 
of the NLO expansion coefficient in the crossover region. 

\begin{figure*}[t]
\begin{center}
\includegraphics[width=78mm]{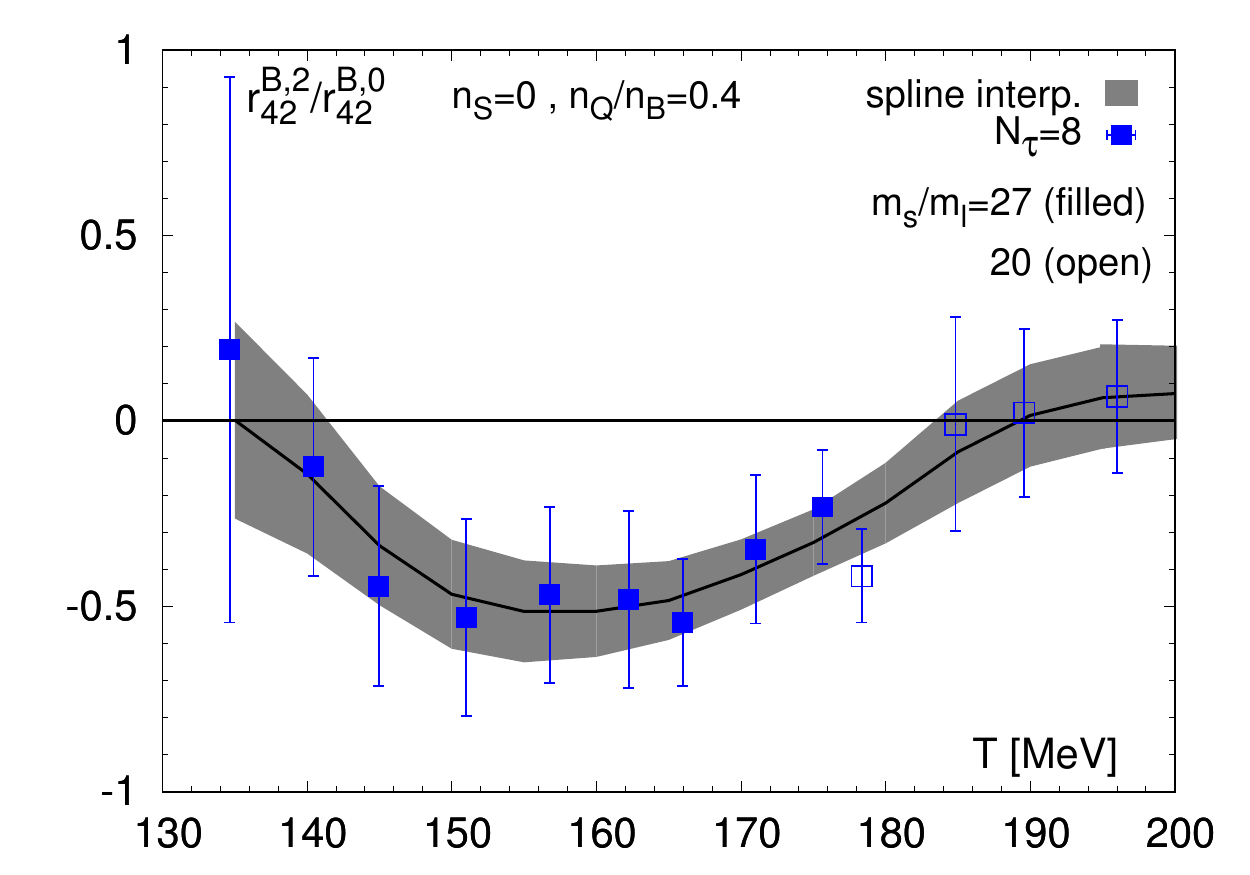}
\includegraphics[width=78mm]{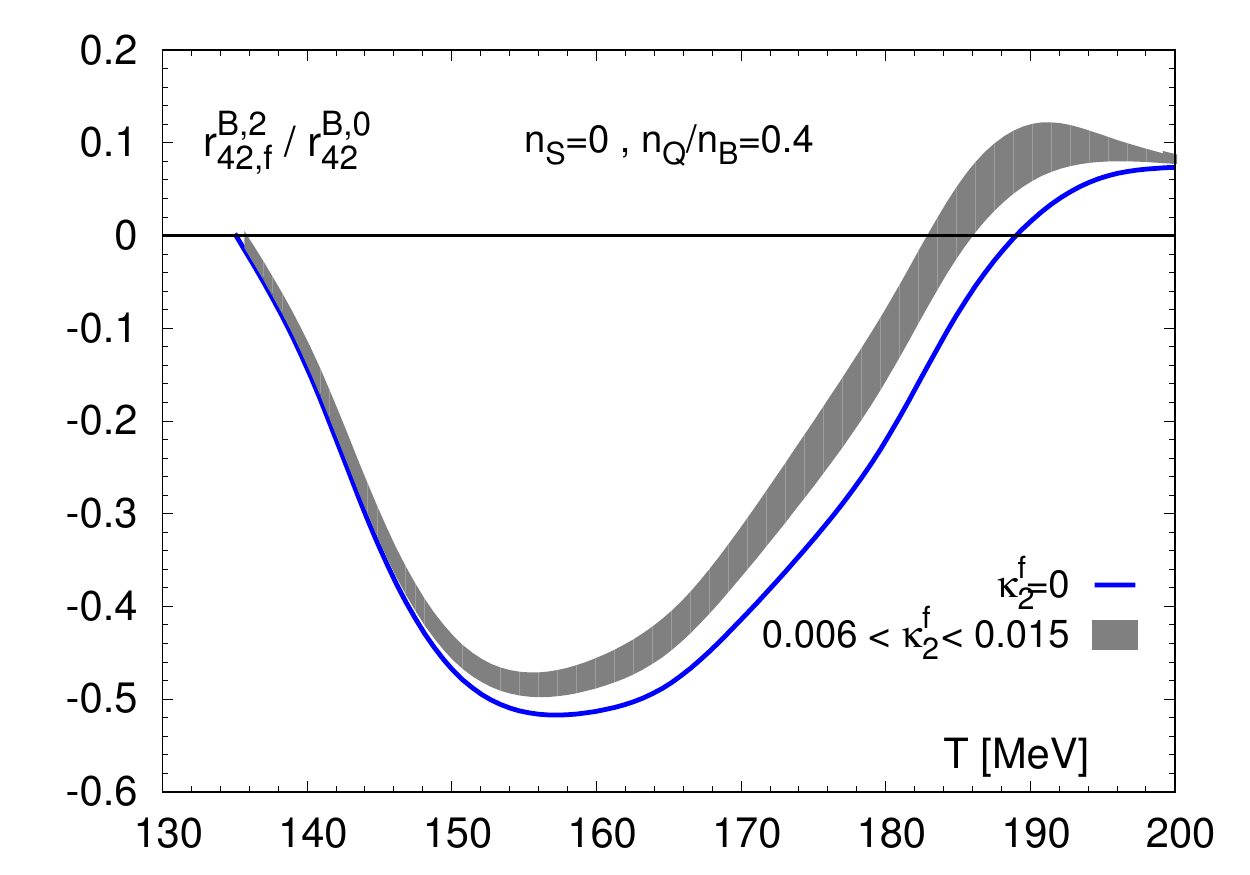}
\caption{{\it Left:} The ratio of NLO and LO expansion coefficients,
$r_{42}^{B,2}/r_{42}^{B,0}$ versus temperature for
a strangeness neutral system, $n_S=0$, with electric charge to baryon number
ratio $n_Q/n_B=0.4$. The black line shows the central value which is
identical to the solid line for the $\kappa_2^f=0$ case shown in the right and
figure.
{\it Right:} The ratio $r_{42,f}^{B,2}/r_{42}^{B,0}$, which gives the
ratio of NLO and LO expansion coefficients evaluated on a line in
the $T$-$\mu_B$ plane as defined in Eq.~\ref{Tf}. The band shows the shift
of this ratio (central values only) resulting from a variation of
$\kappa_2^f$ in the indicated interval.
}
\label{fig:R31B2muS}
\end{center}
\end{figure*}

\begin{figure*}[t]
\begin{center}
\includegraphics[width=78mm]{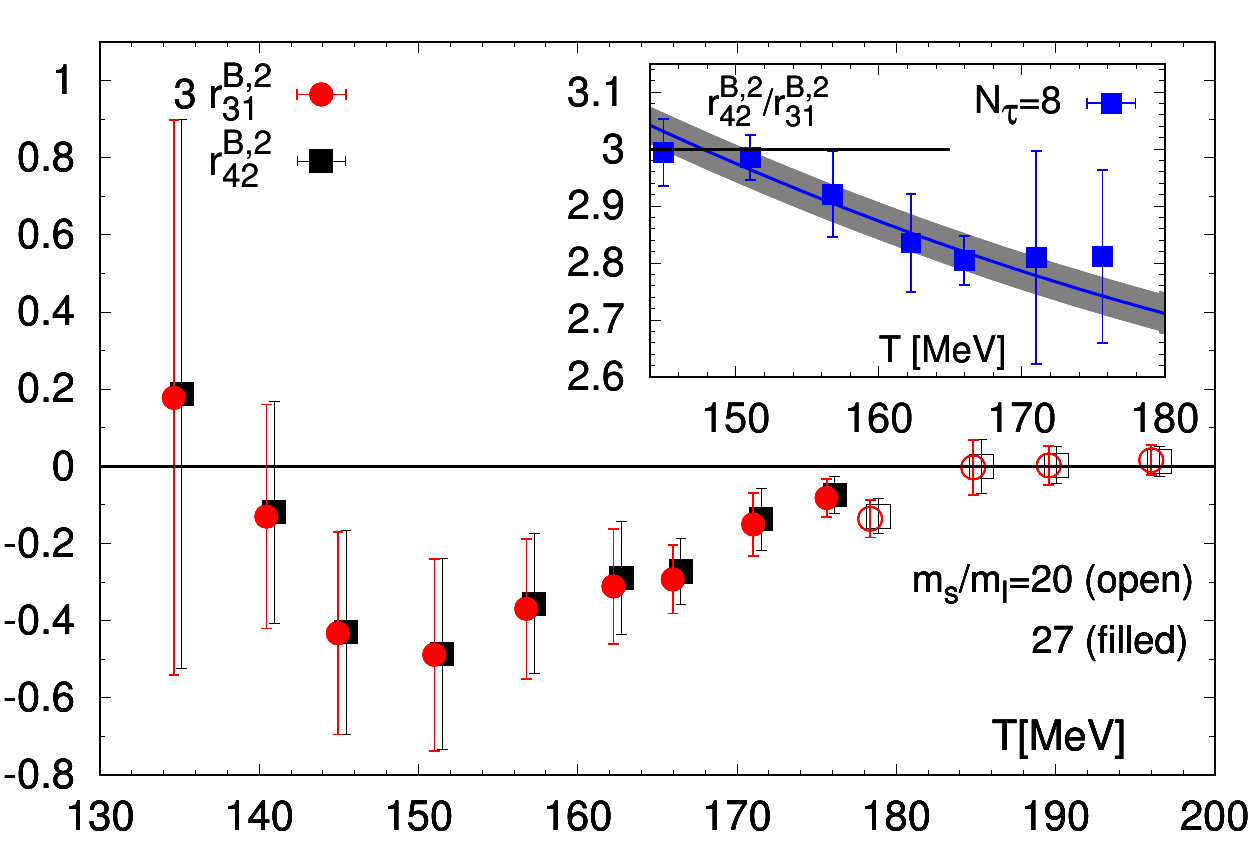}
\includegraphics[width=85mm]{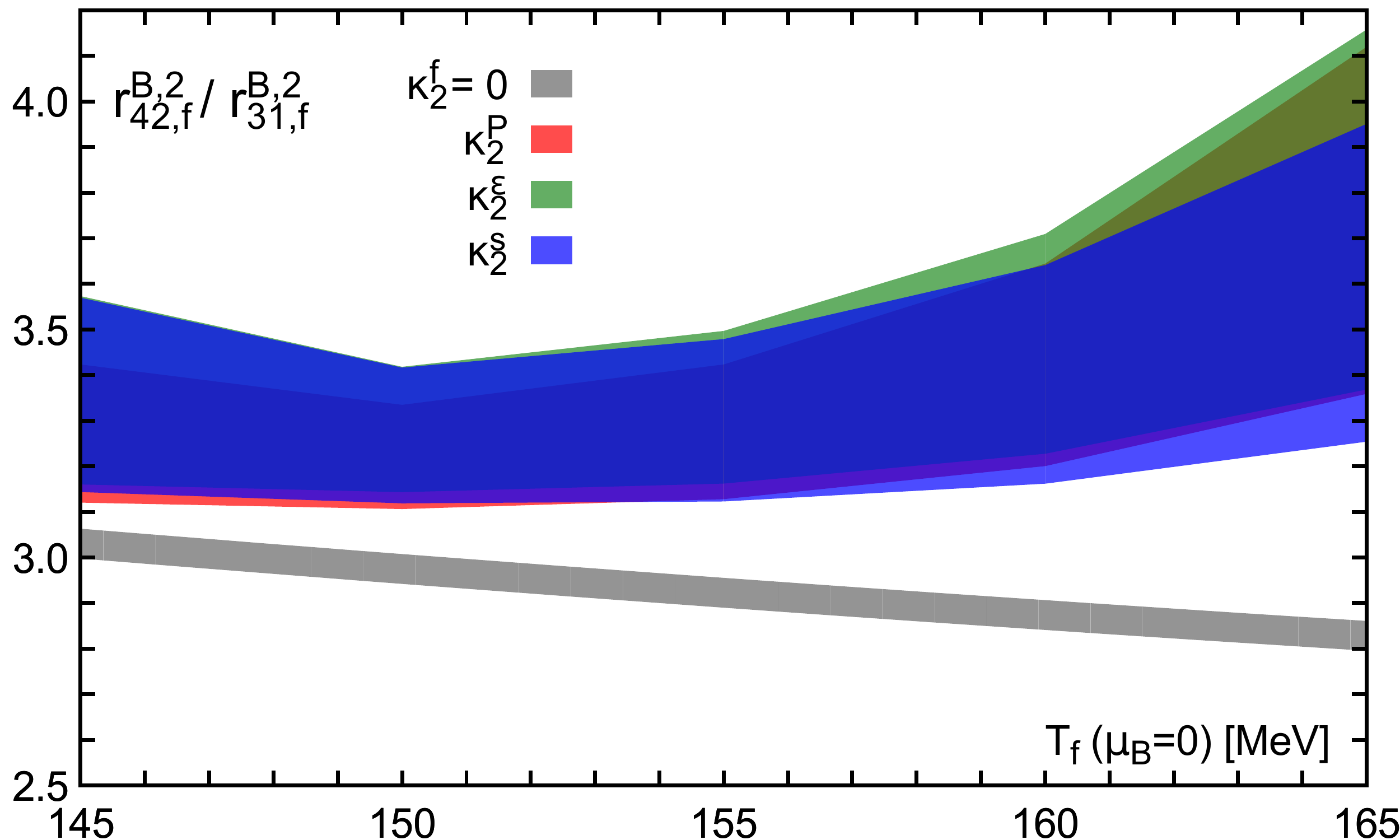}
\caption{{\it Left:} The NLO expansion coefficient for the kurtosis ratio,
$r_{42}^{B,2}$, and three times the NLO expansion coefficient
for the skewness ratio, $r_{31}^{B,2}$.
The inset shows the ratio of the NLO expansion
coefficients, $r_{42}^{B,2}/r_{31}^{B,2}$, in temperature range
where jackknife estimators for this ratio are stable.
{\it Right:}
Ratio of NLO expansion coefficients of the skewness and kurtosis
ratios on lines of constant physics defined by pressure, energy
density and entropy density, respectively. Also shown is the result for
vanishing curvature coefficient ($\kappa_2^f=0$).
Both figures show results for a strangeness neutral system, $n_S=0$, 
with electric charge to baryon number ratio $n_Q/n_B=0.4$.
}
\label{fig:R31_42dif}
\end{center}
\end{figure*}

\subsection{NLO expansion coefficients of \boldmath$R_{42}^B$ and $R_{31}^B$}
The ratio of NLO and LO expansion
coefficients for the kurtosis ratio $R_{42}^B$ is shown in 
Fig.~\ref{fig:R31B2muS}. The left hand figure shows results
for expansion coefficients in the Taylor series evaluated at fixed
temperature. Here only the high statistics lattice QCD results obtained on 
lattices with temporal extent $N_\tau=8$ are shown for the strangeness
neutral case with $n_Q/n_B=0.4$. The band gives a spline interpolation of 
the numerical results. The central line of this interpolation is given as
a black line. 
Although statistical errors are large for these
expansion coefficients, which receive contributions from many
sixth order cumulants, it is apparent that they are negative 
for temperatures $145~{\rm MeV}\lsim T \lsim 175~{\rm MeV}$.

Similar to what has been shown in Fig.~\ref{fig:R12B3} we show 
in the right hand part of Fig.~\ref{fig:R31B2muS} the influence of a 
non-vanishing curvature coefficient, $\kappa_2^f$, on the NLO
expansion coefficients for $R_{42}^B\equiv \kappa_B \sigma_B^2$.
Also in this case the contribution arising from a non-vanishing 
$\kappa_2^f$ is small.
Compared to the LO contribution, however,
the NLO correction to $R_{42}^B$
is large. In the temperature range of interest for a discussion
of freeze-out conditions in heavy-ion collisions, 
$145~{\rm MeV}\lsim T\lsim 165~{\rm MeV}$, the magnitude of 
$r_{42,f}^{B,2}$ varies between 35\% and 50\% of $r_{42}^{B,0}$.

The above observations also hold for the NLO corrections to the 
skewness ratio $R_{31}^B$.
We show a comparison of $r_{42}^{B,2}$ and three times $r_{31}^{B,2}$
in Fig.~\ref{fig:R31_42dif}~(left). Obviously, despite of the large 
statistical errors, the central values of these 
observables match quite well. This hints at a strong correlation
between these two NLO expansion coefficients and allows to 
determine their ratio to much better accuracy than the individual values
would suggest. Nonetheless the jackknife analysis of the ratio
$r_{42}^{B,2}/r_{31}^{B,2}$ still is difficult at low and high
temperature where both observables are compatible with zero within errors.
However, in the temperature 
interval $145~{\rm MeV} < T < 175~{\rm MeV}$ these expansion coefficients
are clearly negative and errors are sufficiently small to determine
the ratio $r_{42}^{B,2}/r_{31}^{B,2}$ reliably. This is shown in the
inset of Fig.~\ref{fig:R31_42dif}~(left). As expected we find that also
in the strangeness neutral case the ratio of expansion coefficients
is close to three, as it is the case for $\mu_Q=\mu_S=0$ 
(see Eq.~\ref{R-R42B20}). 
The ratio
has the tendency to drop with increasing temperature, suggesting that
it will approach the ideal gas value at high temperature\footnote{In the
infinite temperature limit cumulants approach the ideal gas limit. 
For the ratio of NLO expansion coefficients one finds in this limit,
$r_{42}^{B,2}/r_{31}^{B,2}=1.98$ for the strangeness neutral case, $n_S=0$, 
with $n_Q/n_B=0.4$.}.

Using the temperature dependent curvature coefficients $\kappa_2^f$
we can determine the correction to NLO expansion coefficients
of the skewness ($R_{31}^B$) and kurtosis ($R_{42}^B$) ratios, which
arises from a $\mu_B$-dependent freeze-out temperature.
For $\kappa_2^f=0$ the ratio, 
$r_{42}^{B,2}/r_{31}^{B,2}$, is shown as an inset in  
Fig.~\ref{fig:R31_42dif}~(left) together with a quadratic fit.
This ratio drops from $3.03(4)$ at $T=145$~MeV to $2.83(4)$ at $T=165$~MeV.
For $\kappa_2^f>0$ the ratio $r_{42,f}^{B,2}/r_{31,f}^{B,2}$ will
be larger than these values. 
This can be seen from the fact that for $\kappa_2^f>0$
the NLO coefficients $r_{42}^{B,2}$ and $r_{31}^{B,2}$
are shifted by almost the same positive constant (the
temperature derivatives of $r_{42}^{B,0}$ and $r_{31}^{B,0}$
are negative and very similar in magnitude), and in the
temperature range of interest both $r_{42}^{B,2}$ and $r_{31}^{B,2}$
are negative with $r_{42}^{B,2}\sim 3 r_{31}^{B,2}$.
As these derivatives are small
for $T\lsim 145$~MeV, and are largest for
$T\simeq 165$~MeV we thus expect the difference between the ratios
$r_{42,f}^{B,2}/r_{31,f}^{B,2}$ evaluated for $\kappa_2^f=0$ and 
$\kappa_2^f>0$ to rise when increasing the temperature towards the upper 
end of the crossover region. 
This is apparent from the results shown
in  Fig.~\ref{fig:R31_42dif}~(right).
Taking into account current uncertainties on the coefficients $\kappa_2^f$ 
we find in the 
temperature range $145~{\rm MeV} \le T \le 165~{\rm MeV}$,
\begin{equation}
\frac{r_{42,f}^{B,2}}{r_{31,f}^{B,2}} = 3.1 \; -\;  4.1 \; .
\label{r42r31}
\end{equation}

\begin{figure*}[t]
\begin{center}
\includegraphics[width=85mm]{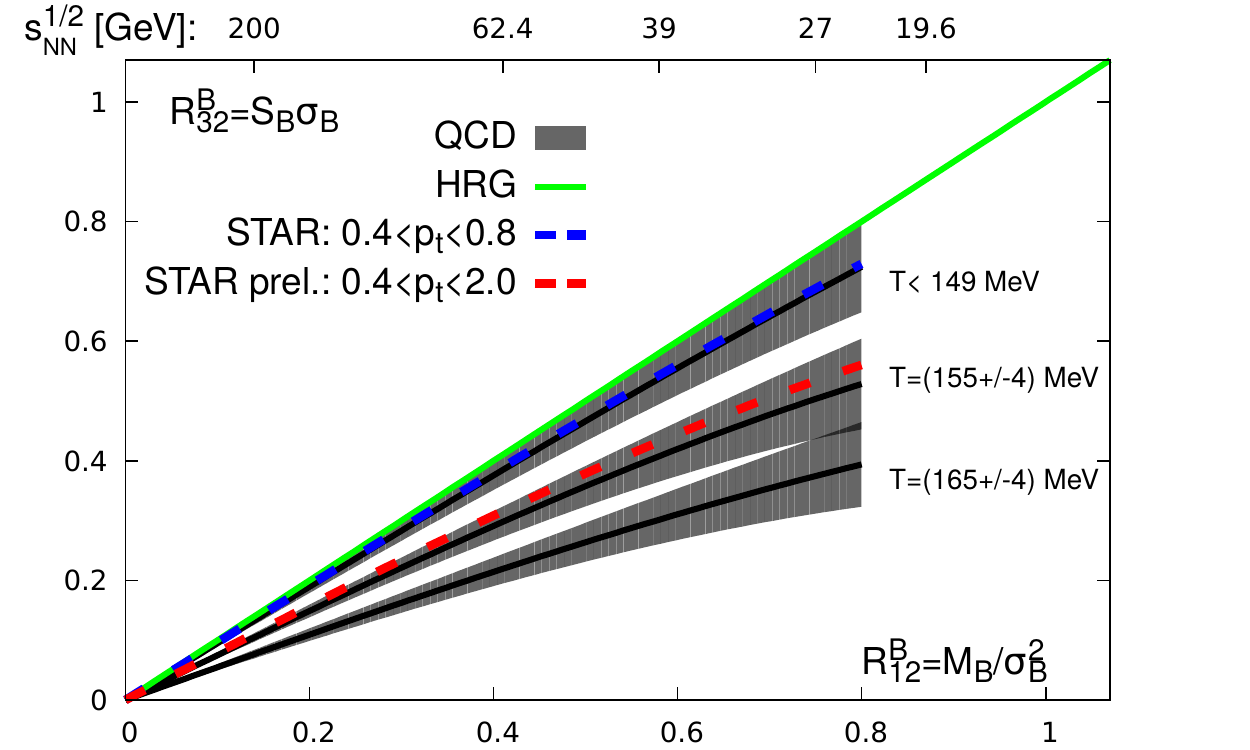}
\includegraphics[width=85mm]{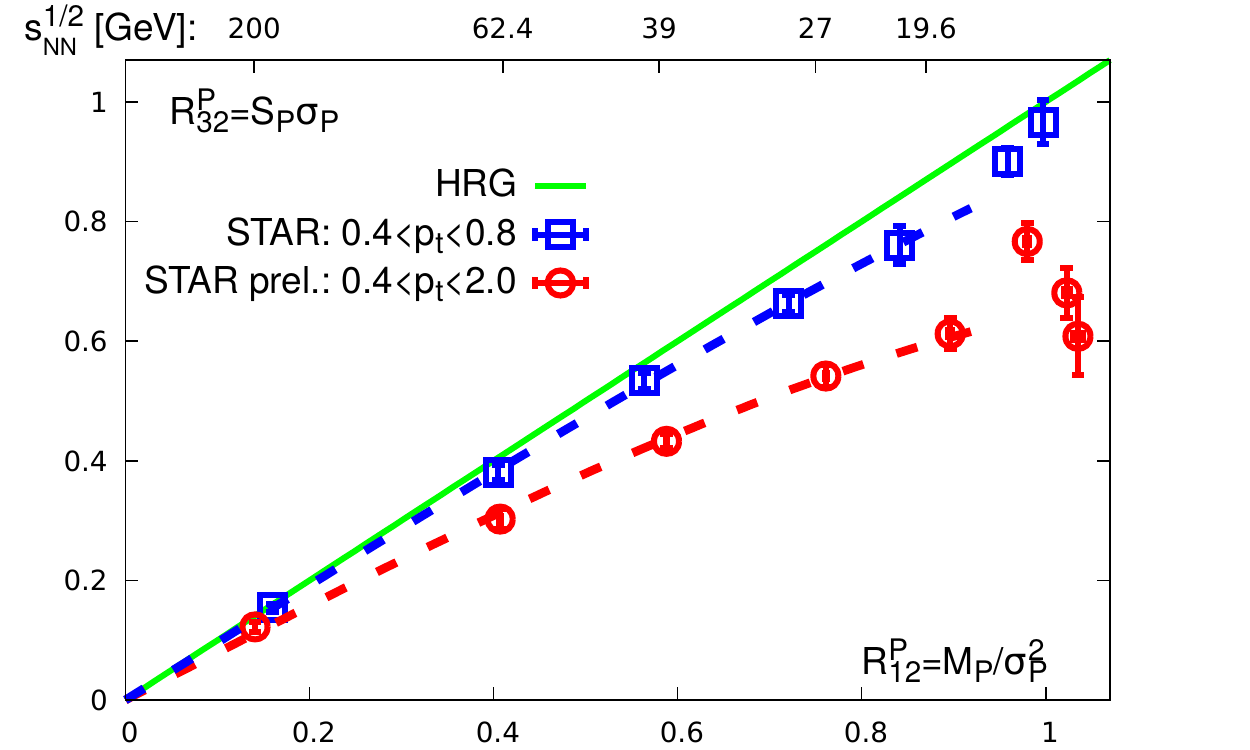}
\caption{{\it Left:}
Next-to-leading order lattice QCD result for 
$S_B\sigma_B$ versus $M_B/\sigma_B^2$.
Bands reflect the statistical
errors of the lattice QCD calculation. The dashed lines show the
fits to the experimental data sets shown in the right hand figure.
{\it Right:} Experimental data on $S_P\sigma_P$ versus $M_P/\sigma_P^2$.
Shown are data obtained by the STAR collaboration
in two different transverse momentum intervals and for different
values of the beam energy, $\sqrt{s_{_{NN}}}/{\rm GeV} = 200,\ 62.4,\
39,\ 27,\ 19.6,\ 11.5,\ 7.7$. For the preliminary data in the
larger $p_t$ range an additional data point is shown at $14.5$~GeV
\cite{Thader:2016gpa}. The dashed lines show cubic fits,
$a R_{12}^P + b \left( R_{12}^P\right)^3$, for $\sqrt{s_{_{NN}}} \ge 19.6$~GeV.
}
\label{fig:R12compare}
\end{center}
\end{figure*}

\section{Comparing NLO lattice QCD calculations with experimental data}
\label{sec:compare}

Qualitative features of the NLO expansions for ratios of cumulants of net 
baryon-number
fluctuations, derived in the previous sections, may be confronted
with experimental results on cumulant ratios of net proton-number
fluctuations. 
Of course, as pointed out in the introduction,
one cannot directly compare the experimental data on net proton-number
fluctuations with those for net baryon-number fluctuations.
In particular, the systematic differences between the two sets of data 
\cite{STARp08,STARp20,Thader:2016gpa}
taken in different transverse momentum intervals, as well as the known 
sensitivity
of the data on acceptance cuts, indicate that these systematic
effects need to be taken care of, e.g.  by implementing them in realistic
hydrodynamic and transport calculations, before a quantitative comparison
becomes possible. A recent study, for instance, suggests that effects of
volume fluctuations are small and also effects arising from
hadronic scattering and resonance decays have little influence
on the ratios of net proton-number cumulants \cite{Li:2017via}.

Since experimentally measured cumulants of net proton-number fluctuations 
can be different from the cumulants of net baryon-number fluctuations
\cite{Kitazawa:2012at}, a direct comparison between the two is subject to 
systematic errors. However, as we will see,
qualitative trends, visible in the experimental data at beam energies
$\sqrt{s_{_{NN}}}\ge 19.6$~GeV, agree well with QCD results on cumulant
ratios and their dependence on the baryon chemical potential. They are
consistent with a freeze-out temperature close to the QCD transition
temperature.

For the comparison with experimental data we will use the expansion of 
higher order cumulant ratios on lines of constant physics in terms of 
$R_{12}^B$ as given in Eqs.~\ref{R31vsR12} to \ref{R32vsR12}. 
This allows to compare lattice QCD calculations with experimental data
without prior determination of the chemical potential, $\mu_B$.

\begin{figure}[htb]
\begin{center}
\includegraphics[width=78mm]{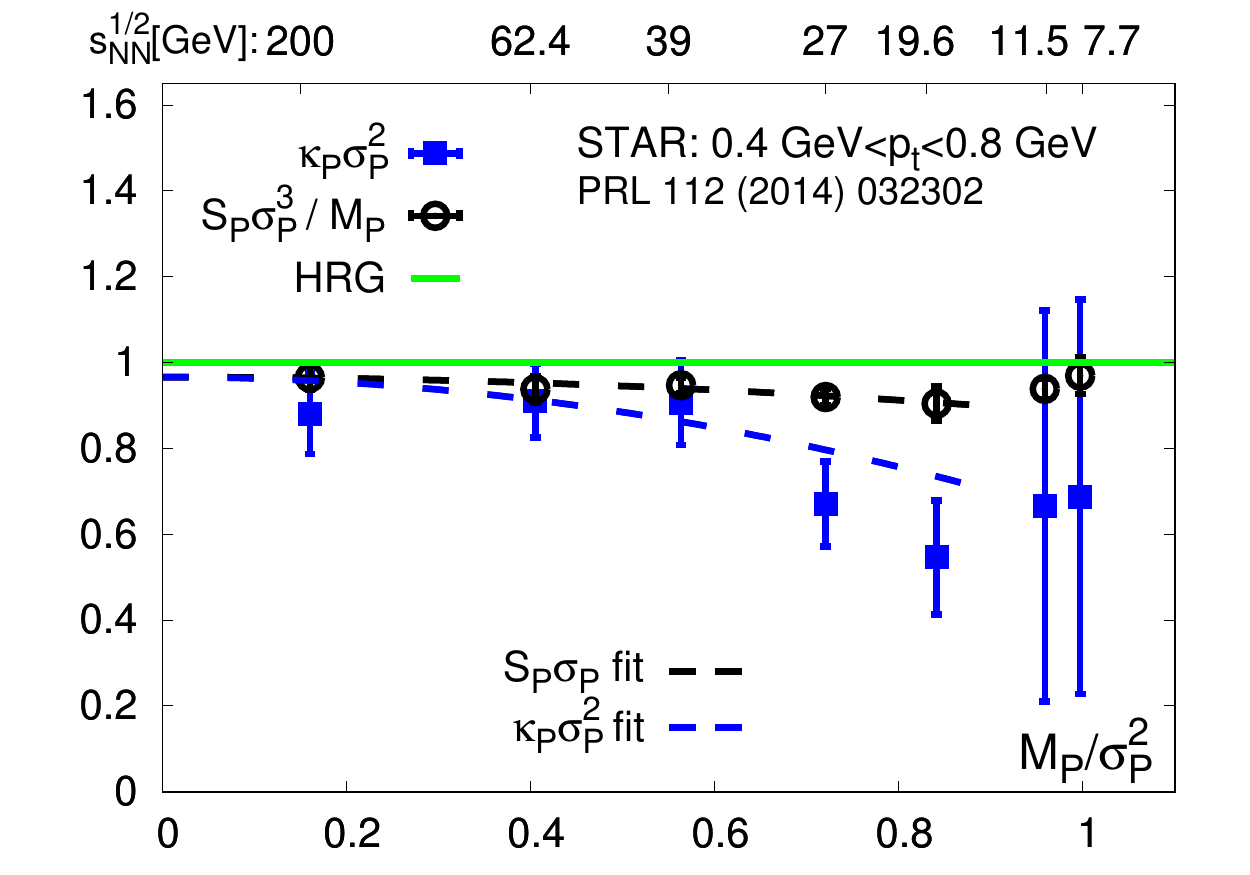}
\includegraphics[width=78mm]{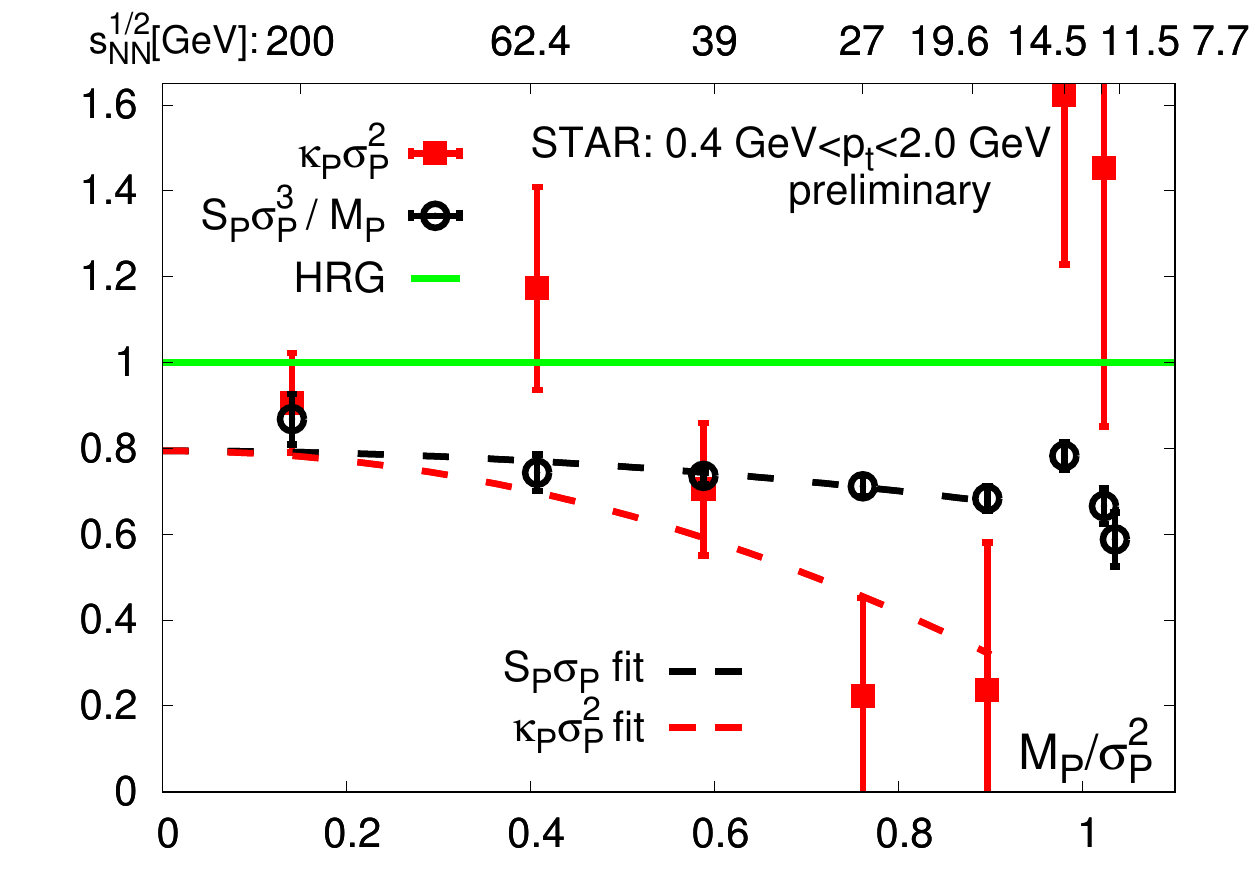}
\caption{Skewness and kurtosis ratios for net proton-number distributions
defined in Eq.~\ref{cumulants} versus mean net proton-number divided by
variance. Data points are for different values of the beam energies as
specified in Fig.~\protect\ref{fig:R12compare}.
In the bottom figure a data point for $\kappa_P\sigma_P^2$ at the
lowest beam energy $\sqrt{s_{_{NN}}}=7.7$~GeV is not shown. See text
for a discussion of the fits.
}
\label{fig:Sk}
\end{center}
\end{figure}

\begin{figure}[t]
\begin{center}
\hspace*{-0.2cm}\includegraphics[width=85mm]{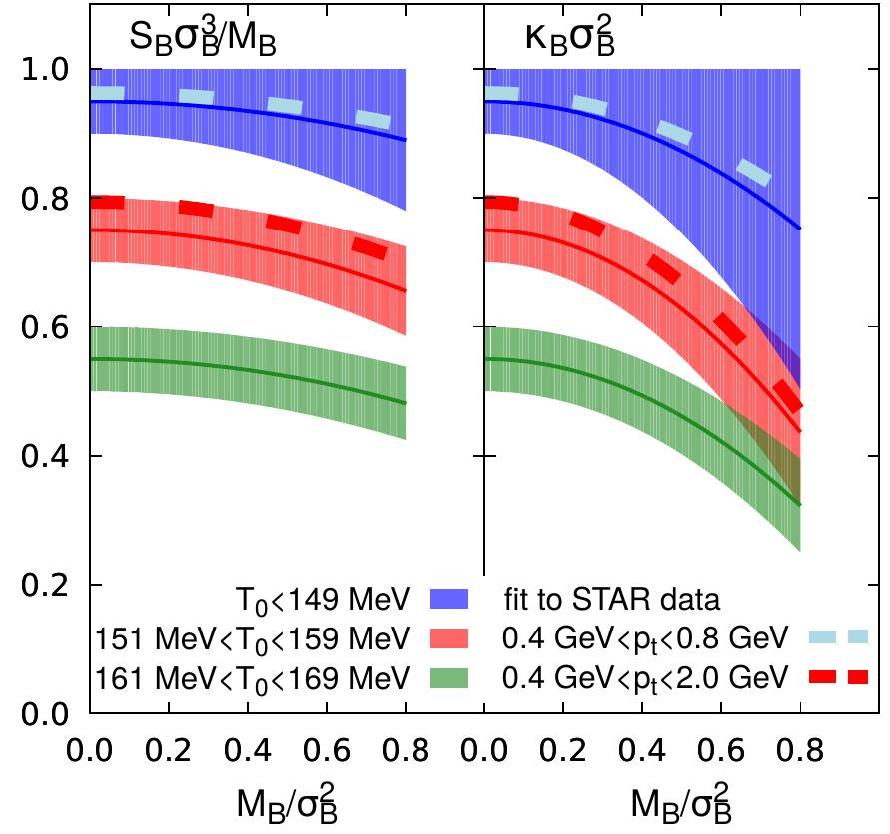}
\caption{NLO expansion of the skewness (left) and kurtosis (right)
ratios $S_B\sigma_B^3/M_B$ and $\kappa_B\sigma_B^2$, respectively. 
Shown are results in three 
temperature ranges covering the crossover region of the QCD transition
at vanishing baryon chemical potential. Dashed lines show the fits to
experimental results for the corresponding skewness and kurtosis ratios
of net proton-number fluctuations. These fits are also shown in
Fig.~\ref{fig:Sk}
}
\label{fig:ratios}
\end{center}
\end{figure}

In Fig.~\ref{fig:R12compare}~(left) we show the skewness ratio 
$R_{32}^B=S_B\sigma_B$ calculated in a NLO Taylor series. Results are
shown as function of $R_{12}^B$ for three temperature intervals.
These temperature intervals have been fixed, somewhat arbitrarily,
by choosing three intervals for the slope parameter, $r_{31}^{B,0}$, i.e.
(i) $r_{31}^{B,0} = 0.95(5)$, (ii) $r_{31}^{B,0} = 0.75(5)$,
(iii) $r_{31}^{B,0} =0.55(5)$. These intervals correspond to temperature
intervals centered around the (i) higher, (ii) central and (iii) lower
value of the QCD transition range $T_c= 154(9)$~MeV. Taking into
account the error band on the spline interpolation shown in 
Fig.~\ref{fig:R12X1}~(right), this leads to the error bands and 
$T$-intervals given in Fig.~\ref{fig:R12compare}~(left).

As is obvious from the temperature dependence of $r_{42}^{B,0}$ 
and $r_{31}^{B,0}$, shown in Fig.~\ref{fig:R12X1}~(right), the 
slope of $R_{32}^B$,  which equals $r_{31}^{B,0}$,  
decreases with increasing temperature and the NLO corrections lead
to a bending of the curves away from the simple straight line result,
which also is obtained in a HRG model calculation with non-interacting, 
point-like hadrons. 
The central temperature range  $151~{\rm MeV} \le T\le 159$~MeV, 
corresponding to the central value obtained for the QCD transition 
temperature, also covers the freeze-out temperature range determined
by the ALICE collaboration at the LHC for almost vanishing chemical 
potential, $T_f= 156(2)$~MeV \cite{Floris:2014pta}. This band also
is consistent with cumulant ratio results obtained by the STAR 
collaboration from an analysis of cumulant ratios measured at 
mid-rapidity, $|y|\le 0.5$, including protons
and anti-protons with transverse momenta 
$0.4~{\rm GeV} \le p_t\le 2.0~{\rm GeV}$ \cite{STARp20,Thader:2016gpa}. 
These data and the 
corresponding STAR data set in a smaller $p_t$-interval, 
$0.4~{\rm GeV} \le p_t\le 0.8~{\rm GeV}$ \cite{STARp08}  are shown 
in Fig.~\ref{fig:R12compare}~(right). We have fitted these data 
for $\sqrt{s_{_{NN}}}\ge 19.6$~{\rm GeV} using
a cubic ansatz, $R_{32}^P = S_0 R_{12}^P + S_2 \left(  R_{12}^P\right)^3$.
The fits for the two different $p_t$-ranges are also shown in
Fig.~\ref{fig:R12compare}~(left). The data obtained in the larger $p_t$-interval
are consistent with freeze-out in the vicinity of the QCD crossover
temperature, while the data in the smaller $p_t$-interval would be
consistent only with a freeze-out temperature
smaller than $150$~MeV.

The STAR data on $R_{32}^P$ versus $R_{12}^P$, obtained at beam energies
$\sqrt{s_{_{NN}}}\ge 19.6$~{\rm GeV}, deviate from a linear dependence
and show evidence for NLO corrections that are consistent in magnitude 
with those determined in the NLO lattice QCD calculation for $R_{32}^B$. The
data obtained in the large $p_t$-interval are thermodynamically consistent
with freeze-out happening close to the QCD transition temperature as well
as a freeze-out temperature $T_f\simeq 156$~MeV as determined by the 
ALICE collaboration. However, these results disfavor a large freeze-out
temperature, $T_f\simeq 165$~MeV, as determined by the STAR collaboration
at large beam energies \cite{Adamczyk:2017iwn}.

The fact that the slope of $R_{32}^B$ differs from unity and decreases
with increasing temperature is equivalent to stating that the intercept
of the skewness ratio $R_{31}^B=S_B \sigma_B^3/M_B$ at vanishing $R_{12}^B$
is smaller than unity and also decreases with increasing temperature.
As can be seen in Eqs.~\ref{R31vsR12} and \ref{R32vsR12} the LO and NLO
expansion coefficients of $R_{32}^B$ and $R_{31}^B$ are, of course,
identical. The fits shown in Fig.~\ref{fig:R12compare}~(right) thus also 
describe the experimental data on the skewness ratio shown in 
Fig.~\ref{fig:Sk}. 
The corresponding results for the skewness ratio of net baryon-number 
fluctuations
obtained from the NLO lattice QCD calculation are shown as black
bands in Fig.~\ref{fig:ratios} for the three temperature intervals
defined previously. These bands
are simply obtained from those shown in 
Fig.~\ref{fig:R12compare}~(left) by dividing with $R_{12}^B$.

The additional fits for the kurtosis ratio $R_{42}^P$ shown in 
Fig.~\ref{fig:Sk} have been obtained by using the quadratic fit
ansatz, $R_{42}^P = K_{0} +K_{2} (R_{12}^P)^2$, with 
$K_{0}\equiv S_{0}$. I.e. we demand that the skewness and kurtosis 
ratios have identical intercept at $R_{12}^P=0$. These constrained
fits provide a good description of the data with  
$K_{2}= (4\pm 2) S_{2}$.
Both fits, shown as blue and red dashed lines in Fig.~\ref{fig:Sk}, are
also shown in Fig.~\ref{fig:ratios}. The ratio $K_2/S_2$ should be 
compared to the ratio of slope parameters, $r_{42,f}^{B,2}/r_{31,f}^{B,2}$, 
for the corresponding
kurtosis and skewness ratios of net baryon-number fluctuations, which
is given in Eq.~\ref{r42r31}. The trend and even the magnitude of this
ratio agrees well with the experimental data.
The stronger bending of the kurtosis
ratio relative to the skewness ratio of net proton-number fluctuations
observed experimentally thus finds a natural explanation in the NLO
lattice QCD calculation.

The general pattern seen in the STAR data for  
$S_P\sigma_P^3/M_P$ and $\kappa_P\sigma_P^2$ for the two different 
$p_t$-intervals is consistent with what
we have discussed for $S_P\sigma_P$ in connection with 
Fig.~\ref{fig:R12compare}. The data obtained in the large 
$p_t$-interval are
consistent with a freeze-out temperature close to the QCD transition
temperature, while the data obtained in the smaller $p_t$-interval are
thermodynamically consistent only with a small freeze-out temperature,
$T_f < 150$~MeV. A large freeze-out temperature of about $165$~{\rm MeV},
on the other hand, would require that the skewness and kurtosis 
ratios become substantially  smaller even at large beam energies once 
all potential systematic corrections have been taken into account. 

The fit to the data for $R_{31}^P$ obtained in the large $p_t$-interval,
which is shown in Fig.~\ref{fig:Sk}, gives the value
$R_{31}^P=R_{42}^P=0.80(4)$ for the intercept at $R_{12}^P=0$. 
This also is shown as a grey box in 
Fig.~\ref{fig:R12X1}~(right) and constrains the range of
freeze-out temperatures quite well. 
We conclude that
all current data on cumulant ratios, measured by STAR in the 
different transverse momentum intervals,
$0.4~{\rm GeV} < p_t < p_t^{\rm cut}$, are consistent with 
freeze-out temperatures,
\begin{eqnarray} 
T_{0} &\le& 149~{\rm MeV}\;\;\;\;\;\;\;\;\;\; {\rm for}\; p_t^{\rm cut} = 0.8~{\rm GeV}\; ,
\nonumber \\
T_{0} &=& (153 \pm 5)~{\rm MeV}\; {\rm for}\; p_t^{\rm cut} =2.0~{\rm GeV}\; .
\end{eqnarray}
The latter is in excellent agreement with the freeze-out temperature
determined by the ALICE Collaboration from particle yields at
the LHC \cite{Floris:2014pta} but differs significantly from the
freeze-out parameters at large beam energies presented by STAR
\cite{Adamczyk:2017iwn}. Within errors it also is consistent
with the somewhat lower value for $T_f$ extracted in our
analysis of ratios of variances of net electric charge and net baryon-number 
fluctuations \cite{Bazavov:2015zja}.

Nonetheless, as stressed above, this 
observation can only be taken
as a first indication, given the observed dependence of
$S_P\sigma_P^3/M_P$ on the transverse momentum range analyzed as
well the size of rapidity bins entering the analysis.

\section{Summary and Conclusions}

Next-to-leading order calculations of cumulant ratios involving
the skewness and kurtosis of net baryon-number fluctuations
are computationally demanding as they involve many $6^{th}$ order
cumulants of conserved charges fluctuations that are difficult to
calculate and statistically noisy in lattice QCD. Depending
on the temperature value the analysis presented here required 
the generation of up to 7 million time units in RHMC simulations
to control these NLO corrections on lattices of size $32^3\times 8$.

Most of the calculations presented here are not yet extrapolated to the 
continuum limit. They, however, clearly show that qualitative 
features of currently available experimental data on net proton-number 
cumulants can be understood
in terms of equilibrium thermodynamics of QCD. In the range of
applicability, $\mu_B~\lsim~ 200$~MeV, which corresponds to
energies $\sqrt{s_{_{NN}}}~\gsim~ 19$~GeV in the RHIC beam energy scan, 
the QCD based calculations presented here may explain 
\begin{itemize}
\item[-] 
the deviation of $S_P\sigma_P^3/M_P$ from unity,
\item[-]
the coincidence of the skewness ratio $S_P\sigma_P^3/M_P$
and the kurtosis ratio $\kappa_P\sigma_P^2$ for large
RHIC beam energy, $\sqrt{s_{_{NN}}}\gsim 200$~GeV,
\item[-]
the significantly stronger decrease of $\kappa_P\sigma_P^2$, in 
comparison to $S_P\sigma_P^3/M_P$, with decreasing
beam energies down to $\sqrt{s_{_{NN}}}= 19.6$~GeV.
\end{itemize}
We have shown that NLO corrections to the skewness and kurtosis ratios, 
evaluated for the strangeness neutral case, are negative in the entire interval 
$145~{\rm MeV} \lsim T \lsim 175~{\rm MeV}$. This also holds
for the case $\mu_Q=\mu_S=0$. As shown in Eqs.~\ref{R-R31B20} and 
\ref{R-R42B20}
it is evident that 
negative values for the skewness and kurtosis ratios
in this case imply $\chi_6^B/ \chi_4^B\ < \ \chi_4^B/ \chi_2^B$.
In fact, the $6^{th}$ order cumulant of net baryon number fluctuations turns
out to be negative in this temperature range. 

It is conceivable that higher order cumulants will start 
changing sign in an irregular pattern for $T\gsim 145$~MeV,
indicating that the radius of convergence of the Taylor series for 
the QCD pressure is limited by a singularity in the complex plane
(strictly alternating signs of expansion coefficients would correspond to 
a singularity for purely imaginary values of $\mu_B/T$). Such a 
scenario disfavors the location of a critical point in the QCD 
phase diagram for $T\gsim 145$~MeV.

The observation that Taylor series for skewness and kurtosis
of net baryon-number fluctuations closely resemble the corresponding 
experimental results for the net proton-number fluctuations
for $\mu_B \le 200\ {\rm MeV}$ is, thus, consistent
with the analysis of the radius of convergence of 
the Taylor series for the pressure and second order net baryon-number 
cumulants \cite{Bazavov:2017dus}, which lead to the conclusion that a 
critical point at $\mu_B/T\le 2$ is disfavored by current lattice 
QCD calculations.

\vspace*{0.2cm} 
\noindent
{\it Acknowledgements:} 
This work was supported in part through Contract No. DE-SC001270 with the
U.S. Department of Energy, through the Scientific Discovery through Advanced
Computing (SciDAC) program funded by the U.S. Department of Energy, Office of
Science, Advanced Scientific Computing Research and Nuclear Physics,
the DOE Office of Nuclear Physics funded BEST topical collaboration and
the NERSC Exascale Application Program (NESAP).
This work was also performed under the auspices of the U.S. Department of 
Energy by Lawrence Livermore National Laboratory under Contract 
DE-AC52-07NA27344. Furthermore, this work was supported by
the Deutsche Forschungsgemeinschaft (DFG) through the grant CRC-TR 211
''Strong-interaction matter under extreme conditions'',
the grant 05P15PBCAA of the German Bundesministerium f\"ur Bildung und
Forschung, grant 283286 of the European Union, the National Natural 
Science Foundation of China under grant numbers 11535012 and 11521064,
and the Early Career Research Award of the Science and Engineering 
Research Board of the Government of India.
Numerical calculations have been made possible through an
INCITE grant of USQCD, ALCC grants in 2016 and 2017, and PRACE grants
at CINECA, Italy. 
These grants provided access to resources on Titan
at ORNL, BlueGene/Q at ALCF, Cori-II at NERSC and Marconi at CINECA.
Additional numerical calculations have been performed on USQCD GPU and KNL
clusters at JLab and Fermilab, as well as GPU clusters at
Bielefeld University and Paderborn University.
We thank the Lawrence Livermore National Laboratory (LLNL) Multiprogrammatic 
and Institutional Computing program for Grand Challenge allocations and time 
on the LLNL BlueGene/Q supercomputer.
We furthermore acknowledge the support of NVIDIA through the CUDA Research
Center at Bielefeld University.

\appendix

\section{NLO expansion coefficients for \boldmath $\chi_n^B$}
\label{appA}
We give here the next-to-leading order results for 
the expansion coefficients of up to fourth order
cumulants of net baryon-number
fluctuations in the constrained case, where $\mu_Q$ and
$\mu_S$ get replaced by Eq.~\ref{mu_expansion}. The expansions
of even and odd order cumulants in terms of the baryon chemical
potential, $\hmu_B=\mu_B/T$, are given by

\begin{eqnarray}
\chi_{2n}^B (T,\mu_B) &=&\bar{\chi}_{2n}^{B,0} +\frac{1}{2} \bar{\chi}_{2n}^{B,2} 
\hmu_B^2 \;\;\;\;\;\;\;\;\;\; , n=1, 2\;\;\;\;\;\;\;\;
\nonumber \\
\chi_{2n+1}^B (T,\mu_B) &=&\bar{\chi}_{2n+1}^{B,1} \hmu_B +\frac{1}{6} 
\bar{\chi}_{2n+1}^{B,3} \hmu_B^3 \ , n=0, 1 .\;~
\label{chin}
\end{eqnarray}
Here the $k$-th order expansion coefficients $\bar{\chi}_n^{B,k}$ are
functions of temperature and the Taylor expansion coefficients $q_i,\ s_i$
of the electric charge and strangeness chemical potentials 
(see Eq.~\ref{mu_expansion}).
The LO expansion coefficients are given by,
\begin{eqnarray}
\bar{\chi}_1^{B,1} &=& \chi_2^B + s_1 \chi_{11}^{BS} + q_1 \chi_{11}^{BQ} \; ,
\nonumber \\
\bar{\chi}_2^{B,0} &=& \chi_2^B \; ,  \nonumber \\
\bar{\chi}_3^{B,1} &=& \chi_4^B + s_1 \chi_{31}^{BS} + q_1 \chi_{31}^{BQ} \; ,
 \nonumber \\
\bar{\chi}_4^{B,0} &=& \chi_4^B \; .
\label{chiBn0}
\end{eqnarray}
We note that in LO the expansion coefficients for even order cumulants
do not depend on the constraint put on strangeness and electric
charge densities, while the odd order expansion coefficients 
explicitly depend on them.
The NLO expansion coefficients
for odd order cumulants, $\bchi_n^{B,3}$, $n=1, 3$, can be written as,

\begin{eqnarray}
\bchi_{n}^{B,3} &=& 
m_n^{B,3}+ 6 s_3 \chi_{n1}^{BS} + 6 q_3 \chi_{n1}^{BQ}  \; ,
\label{bchinB3}
\end{eqnarray}
with
\begin{eqnarray}
m_1^{B,3} &=&
\chi_{4}^{B}
+ \chi^{BS}_{13} s_1^3
+ \chi^{BQ}_{13} q_1^3
+ 3 \chi^{BS}_{22} s_1^2
+3 \chi^{BQ}_{22} q_1^2
\nonumber \\
&&
+ 3 \chi^{BS}_{31} s_1
+3 \chi^{BQ}_{31} q_1
+6 \chi^{BQS}_{211} q_1 s_1
\nonumber \\
&&
+3 \chi^{BQS}_{121} q_1^2 s_1
+3 \chi^{BQS}_{112} q_1 s_1^2
\; ,
\label{m1B3}
\end{eqnarray}
and
\begin{eqnarray}
m_3^{B,3} &=&
\chi_{6}^{B}
+ \chi^{BS}_{33} s_1^3
+ \chi^{BQ}_{33} q_1^3
+3 \chi^{BS}_{42} s_1^2
+3 \chi^{BQ}_{42} q_1^2
\nonumber \\
&&+3 \chi^{BS}_{51} s_1
+3 \chi^{BQ}_{51} q_1
+6 \chi^{BQS}_{411} q_1 s_1
\nonumber \\
&&+3 \chi^{BQS}_{321} q_1^2 s_1
+3 \chi^{BQS}_{312} q_1 s_1^2
\; .
\label{m3B3}
\end{eqnarray} 
Explicit expressions for the NLO expansion coefficients $q_3,\ s_3$ 
of the electric charge and strangeness chemical potentials, needed
in Eq.~\ref{bchinB3}, have been given in Appendix B of 
Ref.~\cite{Bazavov:2017dus}.
Similarly, the NLO expansion coefficients of even order cumulants,
$\bchi_n^B$, $n=2, 4$ are obtained as
\begin{eqnarray}
\bchi_2^{B,2} &=& \chi_{4}^B + s_1^2 \chi_{22}^{BS} +q_1^2 \chi_{22}^{BQ}+
2 s_1 \chi_{31}^{BS} 
\nonumber  \\
&&+2 q_1 \chi_{31}^{BQ}+ 2 q_1 s_1 \chi_{211}^{BQS} \; ,
\label{bchi2B2}
\\
\bchi_4^{B,2} &=& \chi_{6}^B + s_1^2 \chi_{42}^{BS} +q_1^2 \chi_{42}^{BQ}+
2 s_1 \chi_{51}^{BS} 
\nonumber \\
&&+2 q_1 \chi_{51}^{BQ}+ 2 q_1 s_1 \chi_{411}^{BQS} \; .
\label{bchi4B2}
\end{eqnarray}

This gives for the ratios of LO expansion coefficients
introduced in Eqs.~\ref{R12B}-\ref{R42B},
\begin{eqnarray}
r_{12}^{B,1}=\frac{\bchi_1^{B,1}}{\bchi_2^{B,0}}\; ,\;
r_{31}^{B,0}=\frac{\bchi_3^{B,1}}{\bchi_1^{B,1}}\; ,\;
r_{42}^{B,0}=\frac{\bchi_4^{B,0}}{\bchi_2^{B,0}}\; .
\label{rnm0}
\end{eqnarray}
For the NLO expansion coefficients introduced
in Eqs.~\ref{R12B}-\ref{R42B} one then obtains,
\begin{eqnarray}
\frac{r_{12}^{B,3}}{r_{12}^{B,1}} &=& 
\frac{1}{6} \frac{\bchi_1^{B,3}}{\bchi_1^{B,1}}
-\frac{1}{2} \frac{\bchi_2^{B,2}}{\bchi_2^{B,0}} \; ,
\;\;\;\;\;\;\;
\label{R12coefficient} \\
\frac{r_{31}^{B,2}}{r_{31}^{B,0}}&=&
\frac{1}{6} \left( \frac{\bchi_3^{B,3}}{\bchi_3^{B,1}}
-\frac{\bchi_1^{B,3}}{\bchi_1^{B,1}} \right)  \; ,
\label{R31coefficient} \\
\frac{r_{42}^{B,2}}{r_{42}^{B,0}}&=&
\frac{1}{2} \left( \frac{\bchi_4^{B,2}}{\bchi_4^{B,0}}
-\frac{\bchi_2^{B,2}}{\bchi_2^{B,0}} \right) \; .
\label{R42coefficient}
\end{eqnarray}
The corresponding expansion coefficients in the case $\mu_Q=\mu_S=0$ are 
obtained
from these expressions simply by substituting $\bchi\rightarrow \chi$.

\end{document}